\documentclass[journal]{IEEEtran}
\usepackage{cite}
\usepackage{amsmath,amssymb,amsfonts}
\usepackage{algorithmic}
\usepackage{graphicx}
\usepackage{textcomp}
\usepackage{xcolor}
\usepackage{bm}
\usepackage{hyperref}
\usepackage{balance}

\ifCLASSINFOpdf
\else
\fi
\newcommand{\N}{\mathbb{N}}
\newcommand{\NN}{\mathcal{N}}
\newcommand{\R}{\mathbb{R}}

\newcommand{\PP}{\mathsf{P}}
\newcommand{\X}{\mathcal{X}}
\newcommand{\Y}{\mathcal{Y}}
\newcommand{\U}{\mathcal{U}}

\newcommand{\G}{\mathcal{G}}
\newcommand{\E}{\mathbb{E}}

\newcommand{\F}{\mathcal{F}}
\newcommand{\LL}{\mathcal{L}}
\newcommand{\Prob}{\mathbb{P}}
\newcommand{\indicator}{\mathbf{1}}

\newcommand*\diff{\mathop{}\!\mathrm{d}}

\newcommand{\Brace}[1]{\left\{#1\right\}}
\newcommand{\Paren}[1]{\left(#1\right)}
\newcommand{\Bracket}[1]{\left[#1\right]}

\newcommand{\ii}{\mathrm{i}}

\newcommand{\bmg}{\bm{g}}

\DeclareMathOperator{\st}{s. t.}
\DeclareMathOperator*{\argmax}{arg\,max}
\DeclareMathOperator*{\argmin}{arg\,min}
\DeclareMathOperator{\Bern}{Bern}
\DeclareMathOperator{\Unif}{Unif}
\DeclareMathOperator{\diag}{diag}

\newtheorem{definition}{Definition}
\newtheorem{theorem}{Theorem}
\newtheorem{remark}{Remark}
\newtheorem{lemma}{Lemma}

\newtheorem{corollary}{Corollary}

\allowdisplaybreaks

\begin{document}
%
\title{Incremental Redundancy With ACK/NACK Feedback at a Few Optimal Decoding Times}
%
%
%

\author{Hengjie~Yang,~\IEEEmembership{Member,~IEEE,}
        Recep Can Yavas,~\IEEEmembership{Member,~IEEE,} \\
        Victoria Kostina,~\IEEEmembership{Member,~IEEE,}
        and~Richard~D.~Wesel,~\IEEEmembership{Fellow,~IEEE}
\thanks{This research is supported by National Science Foundation (NSF) grant CCF-1955660. An earlier version of this paper was accepted for presentation at the 2022 IEEE International Symposium on Information Theory \cite{Yang_ISIT2022}.

Hengjie Yang is with Qualcomm Technologies, Inc., San Diego, CA 92121, USA  (e-mail: hengjie.yang@ucla.edu).

Recep Can Yavas is with the Department of Electrical and Computer Engineering, National University of Singapore, Singapore, 119077 (e-mail: ryavas@caltech.edu).

Victoria Kostina is with the Department of Electrical Engineering, California Institute of Technology, Pasadena, CA 91125 USA (e-mail: vkostina@caltech.edu).

Richard D. Wesel is with the Department of Electrical and Computer Engineering, University of California at Los Angeles, Los Angeles, CA 90095 USA (e-mail: wesel@ucla.edu). 

}}

\maketitle

\begin{abstract}
Incremental redundancy with ACK/NACK feedback produces a variable-length stop-feedback (VLSF) code constrained to have $m$ decoding times, with an ACK/NACK feedback to the transmitter at each decoding time.  This paper focuses on the numerical evaluation of the maximal achievable rate of random VLSF codes as a function of $m$ for the binary-input additive white Gaussian noise channel, binary symmetric channel, and binary erasure channel (BEC). Leveraging Edgeworth and Petrov expansions, we develop tight approximations to the tail probability of length-$n$ cumulative information density that are accurate for any blocklength $n$. We reduce Yavas \emph{et al.}'s non-asymptotic achievability bound on VLSF codes with $m$ decoding times to an integer program of minimizing the upper bound on the average blocklength subject to the average error probability, minimum gap, and integer constraints. We develop two distinct methods to solve this program. Numerical evaluations show that Polyanskiy's achievability bound for VLSF codes, which assumes $m = \infty$, can be approached with a small $m$ for all three channels. For BEC, we consider systematic transmission followed by random linear fountain coding. This allows us to obtain a new achievability bound stronger than a previous bound and new VLSF codes whose rate further outperforms Polyanskiy's bound.
\end{abstract}


%
\IEEEpeerreviewmaketitle

\section{Introduction}
%
%
%
%
Practical systems such as 3G and beyond employ hybrid automatic repeat request (HARQ) \cite{Mandelbaum1974,Hagenauer1988,Rowitch2000} and incremental redundancy (together also known as the type-II HARQ) to guarantee high reliability. In information theory, the code produced by the type-II hybrid ARQ is called a variable-length stop-feedback (VLSF) code constrained to have $m$ decoding times $n_1, n_2, \dots, n_m$, with each decoding time producing an ACK or NACK feedback to the transmitter.

A VLSF code assumes $M$ infinite-length codewords that are designed and fixed before transmission, where $M$ denotes the message size. During transmission, a stop-feedback symbol `0' indicates that the decoder is not ready to decode and transmission should continue, whereas a `1' signifies that the decoder is ready to decode and the transmitter must stop. The stop feedback only affects the portion of a codeword being transmitted rather than the codeword symbols. VLSF codes are a special case of the more general class of variable-length feedback (VLF) codes.  In general, VLF codes can use feedback to change the codeword symbols that will be transmitted.

The study of VLSF code dates back to Forney \cite{Forney1968} who constructed a VLSF code with sparse periodic decoding times and analyzed its performance in the error exponent regime. Yamamoto and Itoh \cite{Yamamoto1979} constructed a variable-length feedback (VLF) code with sparse periodic decoding times that achieves Burnashev's optimal error exponent \cite{Burnashev1976}. In the non-asymptotic regime, Polyanskiy \emph{et al.} \cite{Polyanskiy2011} showed that the $\epsilon$-capacity $\frac{C}{1 - \epsilon}$ is achievable for a DMC with capacity $C$ and target error probability $\epsilon$ by means of information density decoders and random VLSF codes where stop feedback is sent after every symbol.

We mention a few previous works on VLSF codes with finite decoding times. In \cite{Kim2015}, Kim \emph{et al.} investigated VLSF codes with $m$ periodic decoding times and derived a lower bound on throughput. To minimize the average blocklength, Vakilinia \emph{et al.} \cite{Vakilinia2016} developed the \emph{sequential differential optimization} (SDO) procedure that produces decoding time $n_{i+1}$ based on the knowledge of $n_{i}$, $n_{i-1}$, and their successful decoding probabilities approximated by a differentiable function. The  SDO procedure in \cite{Vakilinia2016} uses a Gaussian model to approximate the probability of successful decoding at each decoding time. Later, variations of the SDO procedure were developed to improve the Gaussian model accuracy \cite{Wang2017, Wesel2018}. The SDO procedure has been utilized to optimize systems that employ incremental redundancy and hybrid ARQ \cite{Wong2017}, and to code for the binary erasure channel \cite{Heidarzadeh2018, Heidarzadeh2019}. 

However, for a given error probability, the Gaussian model developed in all previous works is an application of the central limit theorem (CLT), which typically requires a sufficiently large blocklength. In the non-asymptotic regime, however, this condition is often missed, rendering the Gaussian model inaccurate. This issue becomes especially prominent for decoding times less than $100$. Hence, a refined approximation to tail probabilities is desired for the SDO procedure. In addition, the SDO procedure studied in all previous works assumes real-valued decoding times and can be seen as the solution to an \emph{unconstrained} minimization of the upper bound on average blocklength. Thus, it fails to consider the inherent gap constraint that two decoding times must be separated by at least one.

In statistics, the \emph{Edgeworth expansion} \cite{Edgeworth1905,Peter1992} and \emph{Petrov expansion} \cite{Petrov1975} have been known as  powerful tools to approximate the distribution of a sum of independent and identically distributed (i.i.d.) random variables. A fascinating feature of these expansions is that they only require the knowledge of higher-order \emph{cumulants} of each individual random variable. We refer the reader to \cite[Chapter 2]{Peter1992} for a detailed introduction to Edgeworth expansion and its applications. While the original Edgeworth expansion considers non-lattice random variables (e.g., any continuous random variable), for lattice random variables, Kolassa \cite{Kolassa_book} provided the continuity-corrected Edgeworth series that can be used to approximate the tail probability. In this paper, we apply these tools to approximate the tail probability of a length-$n$ cumulative information density, a quantity that is crucial in the analysis of random fixed-length or VLF codes \cite{Polyanskiy2010, Polyanskiy2011}. Numerical evaluations show that these approximations remain accurate at blocklengths as short as $20$.

In a recent work \cite{Yavas2021}, Yavas \emph{et al.} developed an achievability bound for VLSF codes with $m$ decoding times for the additive white Gaussian noise (AWGN) channel under maximal power constraint $P$. This achievability bound is predicated upon Polyanskiy's information density decoder. By applying CLT for functions to their achievability bound and optimizing decoding times, Yavas \emph{et al.} showed an asymptotic expansion of their bound on the maximum message size $M^*(l, m, \epsilon, P)$ for a given average blocklength $l$, number of decoding times $m$, target error probability $\epsilon$, and maximal power $P$ \cite[Th. 1]{Yavas2021}. They showed that a slight increase in $m$ can dramatically improve the achievable rate of VLSF codes. Unfortunately, due to the nested logarithm term, the expansion is only defined for very large $l$ or small $m$. Yavas \emph{et al.} only numerically show their approximation for $m\le 4$, $\epsilon = 10^{-3}$, and $l\le 2000$. Yavas \emph{et al.} also demonstrated that the decoding times generated from the SDO procedure will yield the same second-order coding rates as attained by their construction of decoding times.

In this paper, we are mainly concerned with numerical evaluations of the maximal achievable rate of VLSF codes with $m$ decoding times at a given message size $M$ and target error probability $\epsilon$ for classical binary-input channels, including the binary-input AWGN (BI-AWGN) channel, the binary symmetric channel (BSC), and the binary erasure channel (BEC). A key problem is to assess whether approaching Polyanskiy's achievability bound for VLSF codes \cite{Polyanskiy2011}, which assumes $m = \infty$, requires a large number of decoding times for practically interesting target error probability $\epsilon$. 

Building upon Yavas \emph{et al.}'s achievability bound, for a fixed information density threshold $\gamma$, we formulate an integer program of minimizing the upper bound on average blocklength over all decoding times $n_1, n_2, \dots, n_m$ subject to average error probability, minimum gap, and integer constraints. Eventually, minimization of locally minimum upper bounds over all information density thresholds $\gamma$ yields the globally minimum upper bound, and this method is referred to as the \emph{two-step minimization}. We develop two methods to numerically evaluate this integer program: the \emph{gap-constrained SDO procedure} and the \emph{discrete SDO procedure}. The former relies on approximating the tail probability with a monotone, differentiable function. The latter only requires an estimate of tail probability at each integer decoding times but comes with added search complexity.

For a given integer $M'$, we derive an error regime $\epsilon\le\epsilon^*(M')$ in which Polyanskiy's stopping at zero technique does not improve the achievability bound for all message sizes $M\le M'$. In this error regime, numerical evaluations show that Polyanskiy's achievability bound can be approached with a finite and relatively small $m$ for classical binary-input channels, including the BI-AWGN channel, BSC, and BEC.

A particular attention is paid to the BEC in that the decoder has the ability to identify the correct message whenever only one codeword is compatible with the unerased received symbols. Motivated by this key observation, we construct a new random VLSF code by first transmitting the $k$-bit message systematically and then applying the random linear fountain coding (RLFC) \cite{MacKay1998}, \cite[Chapter 50]{MacKay2005}. Specifically, after systematic transmission, both the encoder and decoder select the same nonzero basis vector in $\{0,1\}^k$ according to some common randomness. The encoder produces the transmitted symbol by linearly combining the message bits using the selected basis vector. The decoder, known as the \emph{rank decoder}, keeps track of the rank of the generator matrix associated with unerased received symbols. As soon as the rank equals $k$, the decoder stops transmission and reproduces the transmitted $k$-bit message with zero error using the inverse of the generator matrix. 

The systematic transmission followed by RLFC (ST-RLFC) allows us to develop a new VLSF achievability bound that outperforms the state-of-the-art VLSF achievability bound  developed by Devassy \emph{et al.} \cite[Th. 9]{Devassy2016}. More importantly, our bound reduces the $23.4\%$ backoff from capacity at information length $k = 3$ reported in \cite{Devassy2016}. We show that the gap to capacity at $k = 3$ diminishes to $0$ as erasure probability decreases to $0$. This gives a negative answer to the open problem in \cite{Devassy2016} whether the gap to capacity (or to the converse) at $k = 3$ is fundamental. In fact, the backoff percentage from capacity derived from Devassy \emph{et al.}'s result is independent of the erasure probability and therefore must be loose.
The ST-RLFC scheme facilitates a similar integer program that can be solved with the discrete SDO procedure. Numerical evaluations show that the achievable rate of VLSF codes constructed from ST-RLFC scheme significantly outperforms Polyanskiy's achievability bound. For $16$ decoding times, the achievability bound even outperforms Devassy's bound at small values of $k$.

The remainder of this paper is organized as follows. In Section \ref{sec: preliminaries}, we introduce the notation, classical binary-input channels and information density, the VLSF code with $m$ decoding times, and previously known achievability bounds for VLSF codes. In Section \ref{sec: tight approx}, we present tight approximations to the tail probability of length-$n$ information density. In Section \ref{sec: methodology}, we formulate the integer program, establish the properties for optimal decoding times, and present two methods to solve the integer program: gap-constrained SDO and discrete SDO procedures. In Section \ref{sec: numerical evaluations}, we numerically evaluate the achievability bound for VLSF codes with $m$ decoding times for the BI-AWGN, BSC, and BEC, and compare our results with Polyanskiy's achievability bound. In Section \ref{sec: ST-RLFC}, we present the ST-RLFC scheme, a new VLSF achievability bound for the BEC, and numerical evaluations. Section \ref{sec: conclusion} concludes the paper.

\section{Preliminaries}\label{sec: preliminaries}

\subsection{Notation}
Let $\N = \{0, 1,\dots\}$ be the set of natural numbers. For $i\in\N$, $[i]\triangleq \{1,2,\dots, i\}$. We use $x_i^j$ to denote a sequence $(x_i, x_{i+1}, \dots, x_j)$, $1\le i\le j$. When the context is clear, $x_1^n$ is abbreviated to $x^n$. We use $\log(\cdot)$ and $\ln(\cdot)$ to denote the base-$2$ and natural logarithms, respectively. We denote by $\ii$ the imaginary unit, and by $\indicator_{E}$ the indicator function for an event $E$. We use $\phi(x) = \frac{1}{\sqrt{2\pi}}e^{-\frac{x^2}{2}}$, $\Phi(x) = \int_{-\infty}^x \phi(z)\diff z$, and $Q(x) = \int_{x}^\infty \phi(z)\diff z$ to respectively denote the probability density function (PDF), cumulative distribution function (CDF), and the tail probability of a standard normal $\NN(0, 1)$. For a finite, discrete set $\X$, we use $\Unif(\X)$ to denote the uniform distribution over $\X$. We denote the distribution of a random variable $X$ by $\PP_X$.

\subsection{Classical Binary-Input Channels and Information Density}

In this paper, we investigate three memoryless, binary-input channels: the BI-AWGN channel, the BSC, and the BEC.

A BI-AWGN channel consists of input alphabet $\X = \{-\sqrt{P}, \sqrt{P}\}$, output alphabet $\Y = \R$, and conditional PDF
\begin{align}
  \PP_{Y|X}(y|x) = \phi(y - x),\, x\in\X, y\in\Y,
\end{align}
where $P > 0$ denotes the signal-to-noise ratio (SNR). A BSC$(p)$ consists of binary input and output alphabets $\X = \Y = \{0, 1\}$, and conditional probability
\begin{align}
  \PP_{Y|X}(y|x) = p\indicator_{\Brace{y \ne x}} + (1 - p)\indicator_{\Brace{y = x}},\, x\in\X, y\in\Y,
\end{align} 
where $p\in(0, 1/2)$ is called the crossover probability. A BEC$(p)$ consists of a binary input alphabet $\X = \{0, 1\}$, a ternary output alphabet $\Y = \{0, 1, ?\}$, and conditional probability
\begin{align}
  \PP_{Y|X}(y|x) = p\indicator_{\Brace{y = ?}} + (1 - p)\indicator_{\Brace{y = x}},\, x\in\X, y\in\Y,
\end{align}
where $p\in[0, 1)$ is called the erasure probability. For the three channels described above, $\Unif(\X)$ is the capacity-achieving input distribution. 

The information density of a channel $\PP_{Y^n|X^n}$ under input distribution $\PP_{X^n}$ is defined as
\begin{align}
  \iota(x^n; y^n)\triangleq \log\frac{\PP_{Y^n|X^n}(y^n|x^n) }{\PP_{Y^n}(y^n) },
\end{align}
where $\PP_{Y^n}$ is the marginal of $\PP_{X^n}\PP_{Y^n|X^n}$. If $P_{X^n} = \prod_{i=1}^nP_{X_i}$ and the channel is memoryless, we have
\begin{align}
  \iota(x^n; y^n) = \sum_{i = 1}^n\iota(x_i; y_i).
\end{align}
Furthermore, define
\begin{align}
  C &\triangleq \E_{P^*_XP_{Y|X}}[\iota(X; Y)], \\
  V &\triangleq \E_{P^*_XP_{Y|X}}[\iota^2(X; Y)] - C^2,
\end{align}
as the capacity and the dispersion of the channel, respectively, where $P_X^*$ is the capacity-achieving input distribution that is assumed to be unique.


\subsection{VLSF Codes with Finite Decoding Times}
We consider variable-length coding for a memoryless channel $(\X, \Y, \PP_{Y|X})$ with $m$ stop-feedback opportunities. Below, we formally define such codes.
\begin{definition}\label{def: VLSF code}
An $(l, n_1^m, M, \epsilon)$ VLSF code for memoryless channel $(\X, \Y, \PP_{Y|X})$, where $l > 0$, $n_1^m\in\N^m$ satisfying $n_1 < n_2 < \cdots < n_m$, $M\in\N_+$, and $\epsilon\in(0, 1)$, is defined by:
\begin{itemize}
  \item [1)] A finite alphabet $\U$ and a probability distribution $\PP_U$ on $\U$ defining the common randomness random variable $U$ that is revealed to both the transmitter and the receiver before the start of the transmission.
  \item [2)] A sequence of encoders $f_n: \U\times [M] \to \X$, $n = 1,2,\dots, n_m$, defining the channel inputs
    \begin{align}
      X_n = f_n(U, W),
    \end{align}
    where $W\in[M]$ is the equiprobable message.
  \item [3)] A non-negative integer-valued random stopping time $\tau\in\{n_1, n_2, \dots, n_m\}$ of the filtration generated by $\{U, Y^{n_i}\}_{i=1}^m$ that satisfies the average decoding time constraint
    \begin{align}
      \E[\tau] \le l.
    \end{align}
  \item [4)] $m$ decoding functions $g_{n_i}: \U\times \Y^{n_i}\to [M]$, providing the best estimate of $W$ at time $n_i$, $i\in[m]$. The final decision $\hat{W}$ is computed at time instant $\tau$, i.e., $\hat{W} = g_{\tau}(U, Y^\tau)$ and must satisfy the average error probability constraint
    \begin{align}
      P_e \triangleq \Prob[\hat{W}\ne W] \le \epsilon.
    \end{align}
\end{itemize}
\end{definition}
The rate of an $(l, n_1^m, M, \epsilon)$ VLSF code is defined by 
\begin{align}
    R \triangleq \frac{\log M}{\E[\tau]}.
\end{align}

In Definition \ref{def: VLSF code}, the cardinality $\U$ specifies the number of deterministic codes under consideration to construct the random code in Definition \ref{def: VLSF code}. In \cite[Appendix D]{Yavas_TIT2021}, Yavas \emph{et al.} showed that $|\U|\le2$ suffices.

\subsection{Previous Results on VLSF Codes}

In \cite{Polyanskiy2011}, Polyanskiy \emph{et al.} proved a general achievability bound on $(l, \N, M, \epsilon)$ VLSF codes. Namely, they considered $m = \infty$ and $n_i = i-1$ for $i\in\N_+$.
\begin{theorem}[Th. 3, \cite{Polyanskiy2011}]\label{theorem: 1}
  Fix a scalar $\gamma > 0$ and a memoryless channel $(\X, \Y, \PP_{Y|X})$. Let $X^n$ and $\bar{X}^n$ be independent copies from the same process and let $Y^n$ be the output of the channel when $X^n$ is the input. Define a pair of hitting times
  \begin{align}
    \psi &\triangleq \min\Brace{n\ge 0: \iota(X^n; Y^n)\ge\gamma }, \\
    \bar{\psi} &\triangleq \min\Brace{n\ge 0: \iota(\bar{X}^n; Y^n)\ge\gamma },
  \end{align}
Then, for any $M\in\N$, there exists an $(l,\N, M, \epsilon')$ VLSF code satisfying
\begin{align}
  l &\le \E[\psi], \label{eq: th1 E_tau} \\
  \epsilon' &\le (M - 1)\Prob[\bar{\psi}\le \psi]. \label{eq: th1 P_e}
\end{align}
\end{theorem}
The proof of Theorem \ref{theorem: 1} involves generating $M$ infinite-length VLSF codewords at random and an information density  decoder that seeks the smallest stopping time among $M$ stopping times, one for each message. 

In general, it is still difficult to compute $\E[\psi]$ and $\Prob[\bar{\psi}\le \psi]$. Nonetheless, for memoryless channels with bounded information density $\iota(x; y) < \infty$, Polyanskiy \emph{et al.} proved the following useful relaxations by drawing $X^n$ i.i.d. from the capacity-achieving input distribution $P_{X}^*$:
\begin{align}
  &\E[\psi] \le \frac{\gamma + a_0}{C},  \label{eq: th1_eq1} \\
  &\Prob[\bar{\psi} \le \psi] \le 2^{-\gamma},  \label{eq: th1_eq2}
\end{align}
where $a_0 \triangleq \sup_{x\in\X, y\in\Y}\log\frac{P_{Y|X}(y|x)}{\sum_{x'\in\X}P_X^*(x')P_{Y|X}(y|x')}$. Given a target error probability $\epsilon\in(0, 1)$, by setting $\gamma = \log\frac{M-1}{\epsilon}$ in \eqref{eq: th1_eq1} and \eqref{eq: th1_eq2}, \eqref{eq: th1 E_tau} and \eqref{eq: th1 P_e} are further relaxed to
\begin{align}
  &l \le \frac{\log\frac{M-1}{\epsilon} + a_0}{C}, \label{eq: relaxation on blocklength} \\
  &\epsilon' \le \epsilon. \label{eq: relaxation on error prob}
\end{align}
In this paper, we use \eqref{eq: relaxation on blocklength} and \eqref{eq: relaxation on error prob} to evaluate Theorem \ref{theorem: 1}. We remark that in \eqref{eq: relaxation on blocklength}, the term $a_0$ is not tight, hence it is possible to outperform this bound at a finite number of decoding times.

Following the information density framework and a similar argument as in \cite{Polyanskiy2011}, Yavas \emph{et al.}  established an achievability bound for $(l, n_1^m, M, \epsilon)$ VLSF codes for the AWGN channel under maximal power constraint. With a slight modification of removing the maximal power constraint and the violation of power constraint term in the upper bound on error probability, their result holds for an arbitrary memoryless channel. We quote the modified result as follows.
\begin{theorem}[Th. 3, \cite{Yavas2021}]\label{theorem: 2}
  Fix a constant $\gamma > 0$, integer-valued decoding times $n_1 < n_2 < \cdots < n_m$, and a memoryless channel $(\X, \Y, \PP_{Y|X})$. For any $l > 0$ and $\epsilon \in (0, 1)$, there exists an $(l, n_1^m, M, \epsilon')$ VLSF code with
  \begin{align}
    &l \le n_m + \sum_{i=1}^{m-1}(n_i - n_{i + 1})\Prob\Bigg[\bigcup_{j=1}^i \Brace{\iota(X^{n_j}; Y^{n_j})\ge\gamma } \Bigg], \\
    &\epsilon' \le 1 - \Prob[\iota(X^{n_m}; Y^{n_m})\ge\gamma] + (M - 1)2^{-\gamma},
  \end{align}
  where $\PP_{X^{n_m}}$ is the product of distributions of $m$ subvectors of lengths $n_i - n_{i-1}$, $i\in [m]$, i.e.,
  \begin{align}
    \PP_{X^{n_m}}(x^{n_m}) = \prod_{i=1}^m\PP_{X^{n_i}_{n_{i-1}+1}}\Paren{x_{n_{i-1}+1}^{n_i} }. \label{eq: 20}
  \end{align}
\end{theorem}

The proof of Theorem \ref{theorem: 2} is analogous to that of Theorem \ref{theorem: 1}, with the distinction that $X^{n_m}$ is drawn according to \eqref{eq: 20} rather than i.i.d. from a fixed input distribution. In what follows, we assume that $X^n$ is always drawn i.i.d. according to the capacity-achieving input distribution $P_X^*$ unless otherwise specified. This particular choice clearly meets \eqref{eq: 20}.

For the BEC, the decoder in fact has the ability to identify the correct message whenever only a single codeword is compatible with the unerased channel outputs up to that point. By exploiting this fact and utilizing the RLFC, Devassy \emph{et al.} \cite{Devassy2016} obtained better achievability bound for zero-error VLSF codes whose message size $M$ is a power of $2$.
\begin{theorem}[Th. 9, \cite{Devassy2016}]\label{theorem: BEC achievability}
  For each integer $k\ge 1$, there exists an $(l, \N, 2^k, 0)$ VLSF code for a BEC$(p)$ with
    \begin{align}
        l \le \frac{1}{C}\Paren{k + \sum_{i = 1}^{k-1} \frac{2^i - 1}{2^k - 2^i} }, \label{eq: Devassy bound}
    \end{align}
    where $C = 1 - p$.
\end{theorem}

In Section \ref{sec: ST-RLFC}, we present a new upper bound on $l$ (Theorem \ref{theorem: new BEC achiev bound}) using the ST-RLFC scheme that is tighter than  \eqref{eq: Devassy bound}.


\section{Tight Approximations on $\Prob[\iota(X^n; Y^n)\ge\gamma]$}\label{sec: tight approx}

In this section, we develop tight approximations to the tail probability $\Prob[\iota(X^n; Y^n)\ge\gamma]$. Under our construction that $X^n$ are i.i.d., $\iota(X^n; Y^n)$ is a sum of independent random variables distributed the same as $\iota(X_1; Y_1)$. This fact facilitates the use of Edgeworth expansion, Petrov expansion, or Kolassa's continuity-corrected Edgeworth series, all of which can be seen as refined versions of the CLT. A fascinating feature of these expansions is that they only require the knowledge of higher-order cumulants of $\iota(X_1; Y_1)$.

We follow \cite[Chapter I, \S 2]{Petrov1975} in introducing the cumulant of a random variable, which will play an important role in evaluating Edgeworth and Petrov expansions.
\begin{definition}
  For a random variable $W$ with distribution $\PP_W$, let $\chi_W(t) = \E[e^{\ii tW}]$ be its characteristic function. The $j$th cumulant of $W$, $j\ge1$, is defined by the equality
  \begin{align}
    \kappa_j \triangleq \frac{1}{\ii^j}\Bracket{\frac{\diff^j}{\diff t^j}\ln\chi_W(t) }_{t = 0} . \label{eq: cumulant}
  \end{align}
  The characteristic function $\chi_W(t)$ can be expressed in terms of the exponential of the power series of cumulants,
  \begin{align}
      \chi_W(t) = \exp\Paren{\sum_{j=1}^\infty  \frac{\kappa_j}{j!}(\ii t)^j }.
  \end{align}
  In general, the $j$th cumulant $\kappa_j$ is a homogeneous polynomial in noncentral moments of degree $j$, given by
  \begin{align}
    \kappa_j = j!\sum_{\Brace{k_l}}(-1)^{r-1}(r-1)!\prod_{l=1}^j\frac{1}{k_l!}\Paren{\frac{\E[W^l]}{l!} }^{k_l}, \label{eq: cumulants and moments}
  \end{align}
  where the set $\{k_l\}$ consists of all non-negative solutions to $\sum_{l=1}^j l k_l = j$ and $r = \sum_{l=1}^j k_l$.
\end{definition}

\begin{remark}
  Petrov provided the formula \eqref{eq: cumulants and moments} and suggested an induction method as a proof. Blinnikov and Moessner \cite[Appendix B]{Blinnikov1998} presented a direct proof of \eqref{eq: cumulants and moments} and provided an efficient algorithm to compute the set $\Brace{k_l}$ in \eqref{eq: cumulants and moments}.
\end{remark}

\begin{theorem}[Edgeworth Expansion, Eq. (2.18), \cite{Peter1992}]\label{theorem: 3}
    Let $W_1, W_2, \dots, W_n$ be a sequence of i.i.d. random variables with zero mean and a finite variance $\sigma^2$. Define $G_n(x) \triangleq\Prob[\sum_{i=1}^nW_i \le x\sigma\sqrt{n}]$. Let $\chi_W(t)\triangleq \E[e^{\ii tW}]$ be the characteristic function of $W$ and let $\bm{\kappa}=\{\kappa_i\}_{i=1}^\infty$ be the cumulants of $W$. If $\E[|W|^{s+2}] < \infty$ for some $s\in\N_+$ and $\limsup_{|t|\to\infty}|\chi_W(t)| < 1$ (known as Cram\'er's condition), then,
    \begin{align}
    G_n(x) = \Phi(x) + \phi(x)\sum_{j=1}^{s}n^{-\frac{j}{2}}p_j(x) + o\left(n^{-\frac{s}{2}}\right), \label{eq: order-s Edgeworth expansion}
    \end{align}
    where, letting $\bar{\kappa}_i = \sigma^{-i}\kappa_i$ be the order-$i$ cumulant of the normalized random variable $W/\sigma$, 
    \begin{align}
        &p_j(x) {=} -\sum_{\Brace{k_i}}He_{j+2r-1}(x)\prod_{i=1}^j\frac{1}{k_i!}\Paren{\frac{\bar{\kappa}_{i+2}}{(i+2)!} }^{k_i}, \label{eq: p_j poly.}  \\
        &He_j(x) = j!\sum_{k=0}^{\lfloor j/2 \rfloor }\frac{(-1)^kx^{j-2k}}{k!(j-2k)!2^k}, \label{eq: Hermite poly.}
    \end{align}
    and in \eqref{eq: p_j poly.}, the set $\Brace{k_i}$ consists of all non-negative solutions to $\sum_{i=1}^jik_i = j$, $r\triangleq\sum_{i=1}^j k_i$. The polynomial $He_j(x)$ in \eqref{eq: Hermite poly.} is known as the degree-$j$ Hermite polynomial.
\end{theorem}

\begin{IEEEproof}
 We follow the proof in \cite{Peter1992} to derive the Edgeworth expansion. However, we derive explicit formula for the $p_j(x)$ polynomial. See Appendix \ref{appendix: Edgeworth expansion} for the complete proof.
\end{IEEEproof}

\begin{remark}
 In Theorem \ref{theorem: 3}, the Cram\'er's condition holds if the continuous random variable $W$ has a proper density function. If $W$ is a lattice random variable, then Cram\'er's condition in Theorem \ref{theorem: 3} is violated and Theorem \ref{theorem: 3} is no longer applicable. In \cite[Appendix B]{Blinnikov1998}, the authors presented a proof of \eqref{eq: Hermite poly.}.
\end{remark}

We obtain an \emph{order-$s$ Edgeworth expansion} by ignoring the $o(n^{-s/2})$ term in \eqref{eq: order-s Edgeworth expansion}. Meanwhile, \eqref{eq: order-s Edgeworth expansion} suggests that $\lim_{n\to\infty}G_n(x) = \Phi(x)$, which is exactly the CLT.

\begin{theorem}[Petrov Expansion, Chapter VIII, Th. 1, \cite{Petrov1975}]\label{theorem: 4}
    Let $W_1, W_2, \dots, W_n$ be a sequence of i.i.d. random variables with zero mean and a finite variance $\sigma^2$. Define $G_n(x)\triangleq \Prob\Bracket{\sum_{i=1}^nW_i\le x\sigma\sqrt{n} }$. If $x\ge0$, $x = o(\sqrt{n})$, and the moment generating function $\E[e^{tW}]<\infty$ for $|t|<H$ for some $H>0$, then
    \begin{align}
    &G_n(x)=1-Q(x)\exp\Brace{\frac{x^3}{\sqrt{n}}\Lambda\Paren{\frac{x}{\sqrt{n}}} }\Bracket{1 + O\Paren{\frac{x+1}{\sqrt{n} } } }, \\
    &G_n(-x)= Q(x) \exp\Brace{\frac{-x^3}{\sqrt{n}}\Lambda\Paren{\frac{-x}{\sqrt{n}}} }\Bracket{1 + O\Paren{\frac{x+1}{\sqrt{n} } } },
    \end{align}
    where $\Lambda(t) = \sum_{k=0}^\infty a_kt^k$ is called the Cram\'er series. Details on Cram\'er series can be found in the proof of \cite[Chapter VIII, Theorem 2]{Petrov1975}. In particular, Petrov provided the order-$3$ Cram\'er series $\Lambda^{[3]}(t)$
\begin{align}
    &\Lambda^{[3]}(t) \notag\\
    &= \frac{\kappa_3}{6\kappa_2^{3/2}} + \frac{\kappa_4\kappa_2 - 3\kappa_3^2}{24\kappa_2^3}t + \frac{\kappa_5\kappa_2^2 - 10\kappa_4\kappa_3\kappa_2 + 15\kappa_3^3 }{120\kappa_2^{9/2} }t^2, \label{eq: order-2 Cramer series}
\end{align}
where $\{\kappa_i\}_{i=1}^\infty$ denotes the cumulants of random variable $W$.
\end{theorem}

The use of $\kappa_5$ in \eqref{eq: order-2 Cramer series} results in an order-$3$ Petrov expansion in Theorem \ref{theorem: 4}, as can be seen in \eqref{eq: p_j poly.}, where the order of the truncated Edgeworth expansion is determined by the highest order of cumulant minus $2$. Note that at the $0$th order, both Edgeworth and Petrov expansions reduce to $\Psi(x)$.

\begin{remark}
   Both finite-order Edgeworth and Petrov expansions are approximations that are obtained by truncating an infinite series. Edgeworth expansion assumes a constant target probability compared to $n$, whereas Petrov expansion assumes that the target probability decays sub-exponentially to $0$ as $n\to\infty$, defining a moderate deviation sequence in $n$. Therefore, the former performs better in the large $n$ regime, while the latter performs better in small $n$ regime. 
\end{remark}

For lattice random variables, though Theorem \ref{theorem: 3} becomes unavailable, Kolassa \cite{Kolassa_book} provided the \emph{continuity-corrected Edgeworth series} that guarantees the same order of error as usually obtained in an Edgeworth approximation for continuous random variables.
\begin{theorem}[Continuity-Corrected Edgeworth Series, Chapter 3.15, \cite{Kolassa_book}] \label{theorem: corrected Edgeworth}
  Assume that $\{W_i\}_{i=1}^n$ is a sequence of i.i.d. random variables with zero mean, variance $\sigma^2$, and cumulants $\bm{\kappa} = (\kappa_j, j\in\N)$. Suppose that $\{W_i\}_{i=1}^n$ are confined to the lattice $a + u\Delta$, $u\in\N$, $\Delta > 0$, almost surely. Let $Z = \sum_{i=1}^n W_i/(\sigma\sqrt{n})$ and let $F_Z(z)$ be its CDF. Let $\lambda_j^n = \kappa_j - \epsilon_j n^{-1}$ be the adjusted cumulants known as the Sheppard-adjusted cumulants, where $\epsilon_j = (\Delta/\sigma)^jB_j/j$, with $B_j$ being the $j$th Bernoulli number \cite[Chapter 15]{Ireland1990}. Let $\bm{\lambda}^n$ denote the infinite vector $(0, \lambda_2^n, \lambda_3^n,\dots)$. Let 
  \begin{align}
    E_{s}(z; \bm{\kappa}) \triangleq \Phi(z) + \phi(z)\sum_{j=1}^s n^{-\frac{j}{2}} p_j(z)
  \end{align}
  be the order-$s$ Edgeworth expansion evaluated at $z$, with $p_j(\cdot)$ polynomials computed from the first $s+2$ cumulants of $\bm{\kappa}$. Then, 
    \begin{align}
      F_Z(z^+) = E_{s}(z^+; \bm{\lambda}^n) + o\Paren{n^{-\frac{s}{2}} }, \label{eq: corrected Edgeworth}
    \end{align}
    where $z^+ = z + \frac{\Delta/2}{\sigma\sqrt{n}}$ denotes the continuity-corrected lattice point for any lattice point $z$ of $Z$, and the $p_j(\cdot)$ polynomials in $E_{s}(z^+; \bm{\lambda}^n)$ are computed from the first $s+2$ Sheppard-adjusted cumulants in $\bm{\lambda}^n$ rather than $\bm{\kappa}$.
\end{theorem}

Note that for any lattice point $z$, $F_Z(z') = F_Z(z^+)$ for all $z'\in\big[z, z + \frac{\Delta}{\sigma\sqrt{n}}\big)$. Therefore, \eqref{eq: corrected Edgeworth} provides an accurate approximation for the entire real line.

Next, we discuss the approximation to $\Prob[\iota(X^n; Y^n)\ge\gamma]$ for BI-AWGN channel, the BSC, and the BEC. Due to the distinct nature of these channels, we choose different strategies to approximate or exactly evaluate $\Prob[\iota(X^n; Y^n)\ge\gamma]$ for each of these channels. For brevity, denote by $F_{\gamma}(n)$ the function we use to estimate $\Prob[\iota(X^n; Y^n)\ge\gamma]$. The domain of $n$ for $F_{\gamma}(n)$ also depends on the type of channel.

\begin{figure}[t]
\centering
\includegraphics[width=0.48\textwidth]{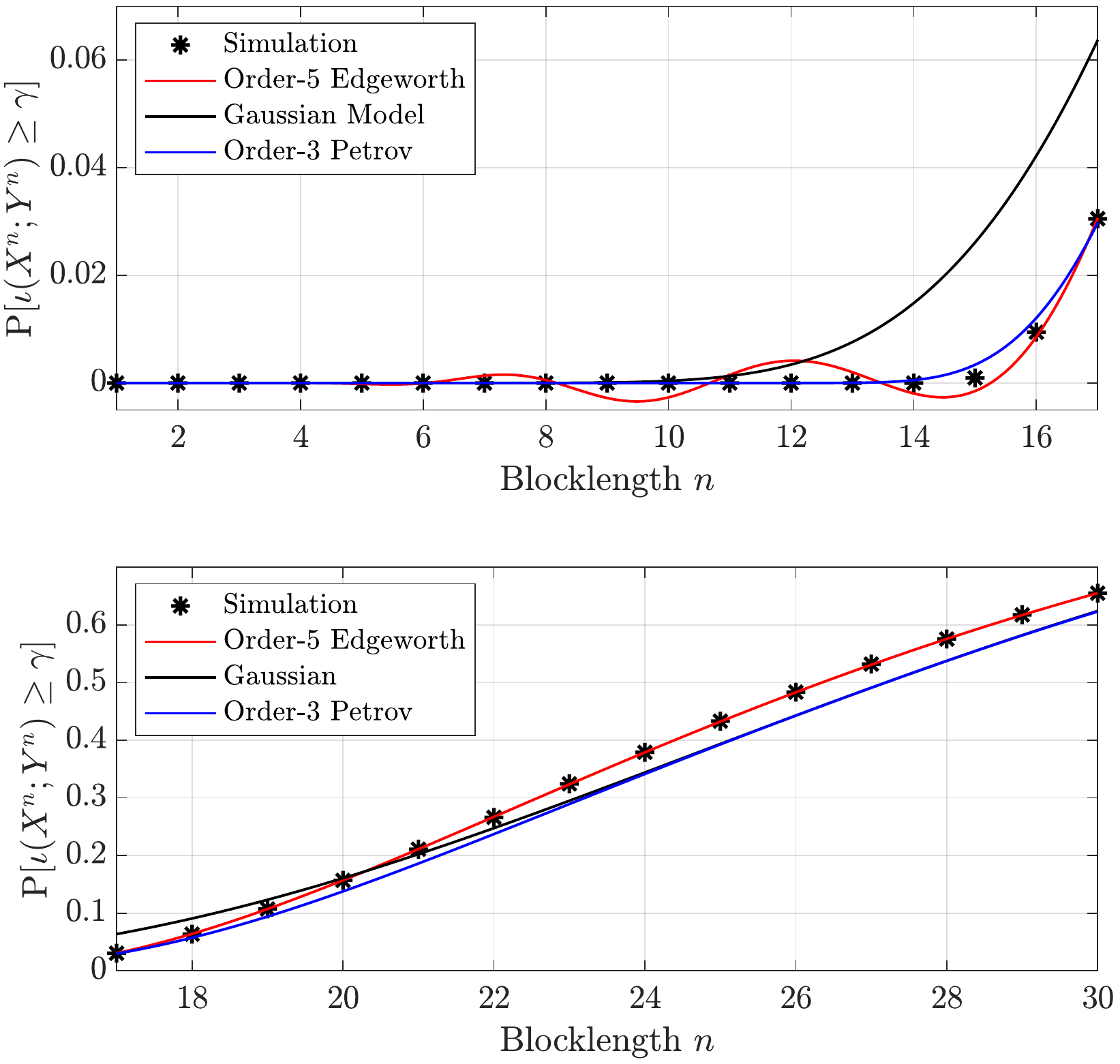}
\caption{Comparison of various approximation models for $\Prob[\iota(X^n; Y^n)\ge \gamma]$ with a fixed $\gamma = 13.62$ for BI-AWGN channel at $0.2$ dB. }
\label{fig: approx models for BI-AWGN}
\end{figure}

\subsection{BI-AWGN Channel}

For the BI-AWGN channel, the information density $\iota(X; Y)$ under $\Unif(\X)$ is given by
\begin{align}
  \iota(X; Y) &= 1 - \log\Paren{1 + e^{-2XY} }.
\end{align}
Clearly, $\iota(X; Y)$ is a continuous random variable with a proper density function. Hence, the order-$s$ Edgeworth expansion in Theorem \ref{theorem: 3} is applicable in this case. In our experimentation, we identify that $s=  5$ meets our desired approximation accuracy at large $n$.

However, a caveat of the order-$s$ Edgeworth expansion is that for small values of $n$, the order-$s$ Edgeworth expansion oscillates around $0$ due to the truncation of an infinite series. To illustrate this issue, Fig. \ref{fig: approx models for BI-AWGN} shows the order-$5$ Edgeworth expansion depicted in solid red curve to approximate $\Prob[\iota(X^n; Y^n)\ge\gamma]$, where $\gamma = 13.62$ and the BI-AWGN channel is at $0.2$ dB. We use Monte Carlo simulation to obtain the tail probability at each time instant. As can be seen, the order-$5$ Edgeworth expansion oscillates around $0$ for $n < 16$. Yet beyond this range, the order-$5$ Edgeworth expansion seamlessly matches the simulated tail probability. 

To circumvent the oscillation issue, we resort to the order-$3$ Petrov expansion in Theorem \ref{theorem: 4} for small $n$ satisfying $n < \gamma/C$. More specifically, the $F_{\gamma}(n)$ we use is given by
\begin{align}
  F_{\gamma}(n) = \begin{cases}
    Q\Paren{x(n)} - \phi(x(n))\sum_{j=1}^5 n^{-\frac{j}{2}}p_j(x(n)),\ n > n^* \\
    Q(x(n))\exp\Brace{\frac{x^3(n)}{\sqrt{n}}\Lambda^{[3]}\left(\frac{x(n)}{\sqrt{n}}\right) },\ 0\le n\le n^*, 
  \end{cases}
  \label{eq: approx function for BI-AWGN}
\end{align}
where $x(n)\triangleq \frac{\gamma - nC}{\sqrt{nV}}$ and $n^* < \gamma/C$ is the largest $n$ value for which two expansions are equal with a common value less than $1/2$.

Fig. \ref{fig: approx models for BI-AWGN} illustrates the order-$3$ Petrov expansion as depicted in solid blue curve. We see that the order-$3$ Petrov expansion provides a good approximation for $n < 16$ yet starts to deviate from the simulated tail probability as $n$ increases. Thus, combining both expansions at switching threshold $n^* = 16.84$ will provide a good approximation over the entire range of blocklength. Fig. \ref{fig: approx models for BI-AWGN} also shows the Gaussian model $Q\Paren{x(n)}$ considered in \cite{Wang2017}, which corresponds to the order-$0$ Edgeworth expansion. As can be seen, the Gaussian model is inaccurate over the entire range of blocklength.

\subsection{BSC}

For the BSC$(p)$, $p\in(0, 1/2)$, the information density $\iota(X; Y)$ under $\Unif(\X)$ is given by
\begin{align}
  \iota(X; Y) &= \log(2 - 2p) - (X\oplus Y)\log\frac{1-p}{p} \\
    &= \log(2 - 2p) - Z\log\frac{1-p}{p}, 
\end{align}
where $Z\sim \Bern(p)$. Hence, $\iota(X; Y)$ is a random walk taking steps $\log(2p)$ and $\log(2-2p)$ with probabilities $p$ and $1 - p$. The length-$n$ information density is thus given by
\begin{align}
  \iota(X^n; Y^n) = n\log(2 - 2p) - \Paren{\log\frac{1-p}{p}}\sum_{i=1}^nZ_i.
\end{align}
The tail probability $\Prob[\iota(X^n; Y^n)\ge\gamma]$ can be computed  from the CDF of the binomial distribution. Hence,
\begin{align}
\Prob[\iota(X^n; Y^n)\ge\gamma] &= \Prob\Bracket{\sum_{i=1}^nZ_i \le \frac{n\log(2 - 2p) - \gamma}{\log\Paren{(1-p)/p }} } \notag\\
    &= \sum_{c = 0}^{\big\lfloor\frac{n\log(2 - 2p) - \gamma}{\log\Paren{(1-p)/p }} \big\rfloor }\binom{n}{c}p^c(1 - p)^{n - c}. \label{eq: CDF of binomial}
\end{align}
Thus, we choose $F_{\gamma}(n) = \Prob[\iota(X^n; Y^n)\ge\gamma]$ which is given by \eqref{eq: CDF of binomial}, where $n\in\N$. Next, we show that for a fixed $\gamma > 0$, $F_{\gamma}(n)$ exhibits a zig-zag shape as $n$ increases.
\begin{theorem}\label{theorem: 6}
  Fix $\gamma > 0$ and $p\in(0, 1/2)$. Define $\alpha_i \triangleq \Big\lceil \frac{\gamma + i\log((1-p)/p) }{\log(2 - 2p)} \Big\rceil$, $i\in\N$. Then, if $n = \alpha_i - 1$, $F_{\gamma}(n) < F_{\gamma}(n + 1)$; if  $n \in [\alpha_i, \alpha_{i+1} - 1)$, $F_{\gamma}(n) > F_{\gamma}(n + 1)$, where $i\in\N$.
\end{theorem}

\begin{IEEEproof}
  First, we show that the interval $[\alpha_i, \alpha_{i+1}-1)$ contains at least one integer. This is because
    \begin{align}
      \alpha_{i+1} - \alpha_i &\ge \frac{\gamma + (i+1)\log\frac{1-p}{p}}{\log(2-2p)} - \left\lceil \frac{\gamma + i\log(\frac{1-p}{p} ) }{\log(2 - 2p)} \right\rceil \notag\\
        & > \frac{\log\frac{1}{p} + \log(1 - p) }{1 + \log(1 - p) } - 1 \notag\\
        & > 1.
    \end{align}
  It follows that $(\alpha_{i+1} - 1) - \alpha_i \ge 1$, implying that the interval $[\alpha_i, \alpha_{i+1}-1)$ contains at least one integer. Next, it suffices to show the result for a fixed $i\in\N$. If $n = \alpha_i - 1$,
    \begin{align}
       F_{\gamma}(n) = \Prob\Bracket{\sum_{j=1}^nZ_j \le i-1 } &< \Prob\Bracket{\sum_{j=1}^nZ_j \le i- Z_{n+1} } \label{eq: increasing} \\
        &= F_{\gamma}(n+1), \notag
     \end{align} 
  where \eqref{eq: increasing} follows from $Z_{n+1}\in\{0, 1\}$. If $n\in[\alpha_i, \alpha_{i+1}-1)$, 
    \begin{align}
      F_{\gamma}(n) = \Prob\Bigg[\sum_{j=1}^nZ_j \le i \Bigg] & > \Prob\Bigg[\sum_{j=1}^{n}Z_j \le i - Z_{n+1} \Bigg] \label{eq: decreasing} \\
        & = F_{\gamma}(n + 1), \notag
    \end{align}
  where \eqref{eq: decreasing} follows from $Z_{n+1}\in\{0, 1\}$. 
\end{IEEEproof}

\begin{figure}[t]
\centering
\includegraphics[width=0.48\textwidth]{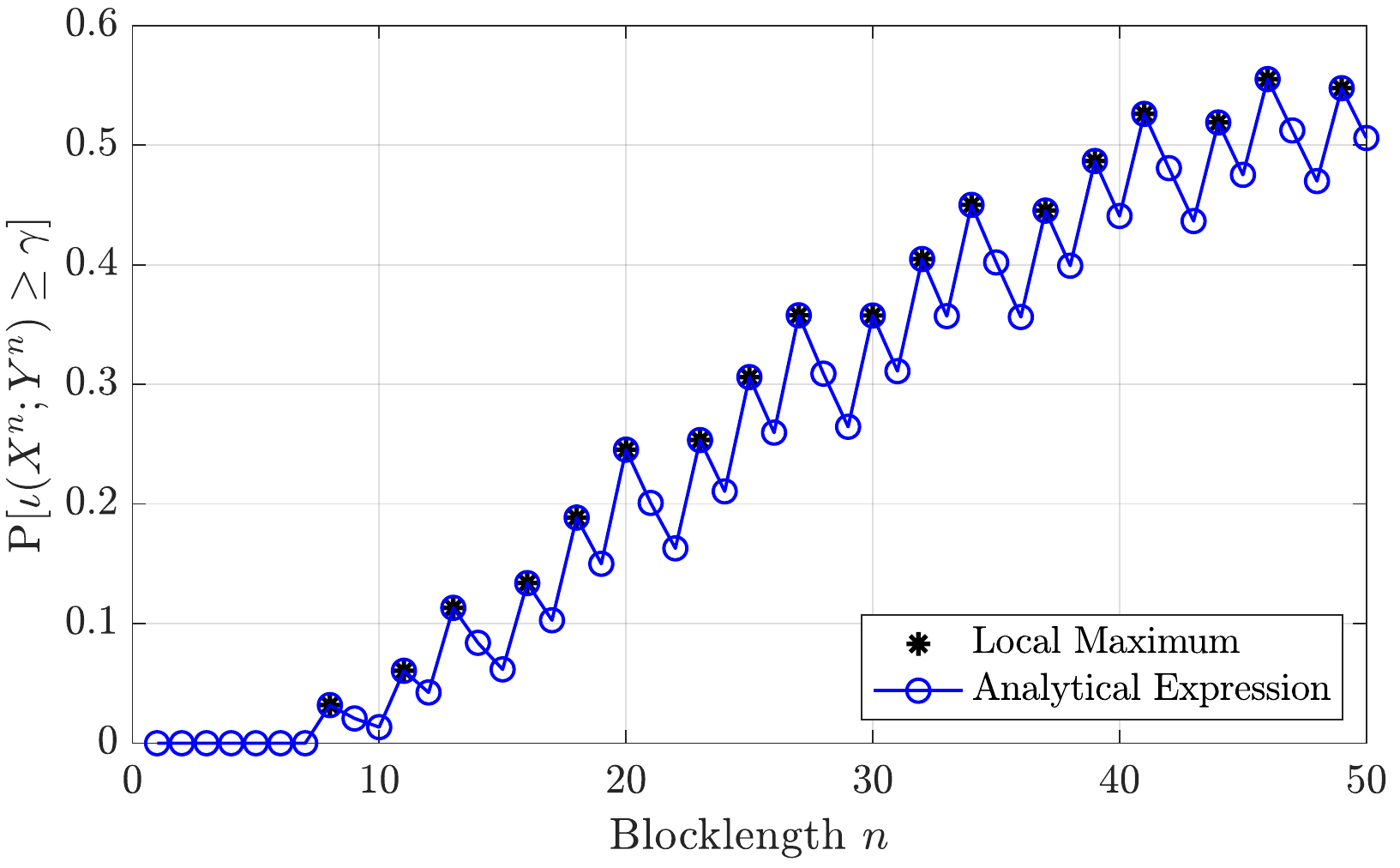}
\caption{Tail probability $\Prob[\iota(X^n; Y^n)\ge \gamma]$ for BSC$(0.35)$ with $\gamma = 3$. }
\label{fig: tail prob for BSC}
\end{figure}

Theorem \ref{theorem: 6} implies that the sequence $\{\alpha_i\}_{i=0}^\infty$ corresponds to the set of local maximizers, whereas the sequence $\{\alpha_i - 1\}_{i=0}^\infty$ corresponds to the set of local minimizers.
As an example, Fig. \ref{fig: tail prob for BSC} shows the tail probability $F_{\gamma}(n)$ as a function of blocklength $n$, which exhibits a zig-zag behavior. The local maximum values correspond to the tail probability at local maximizers $\{\alpha_i\}_{i=0}^\infty$. We see that the tail probabilities at local maximizers may not be a monotonically increasing sequence.

\subsection{BEC}

For BEC$(p)$, $p\in(0, 1)$, the information density $\iota(X; Y)$ under $\Unif(\X)$ is given by
\begin{align}
  \iota(X; Y) = 1 - \indicator_{\Brace{Y \ne X}} = 1 - Z, 
\end{align}
where $Z \sim \Bern(p)$. Thus, the tail probability can be computed from the CDF of binomial distribution.
\begin{align}
  \Prob[\iota(X^n; Y^n)\ge\gamma] &= \Prob\Bracket{\sum_{i=1}^nZ_i \le n - \gamma } \\
    &= \sum_{c = 0}^{\lfloor n -\gamma \rfloor}\binom{n}{p}p^c(1 - p)^{n - c}. \label{eq: tail prob for BEC}
\end{align}
Note that $\iota(X; Y)\in\{0, 1\}$, it follows that $\Prob[\iota(X^n; Y^n)\ge\gamma]$ is a strictly increasing function of $n$. 

\begin{figure}[t]
\centering
\includegraphics[width=0.48\textwidth]{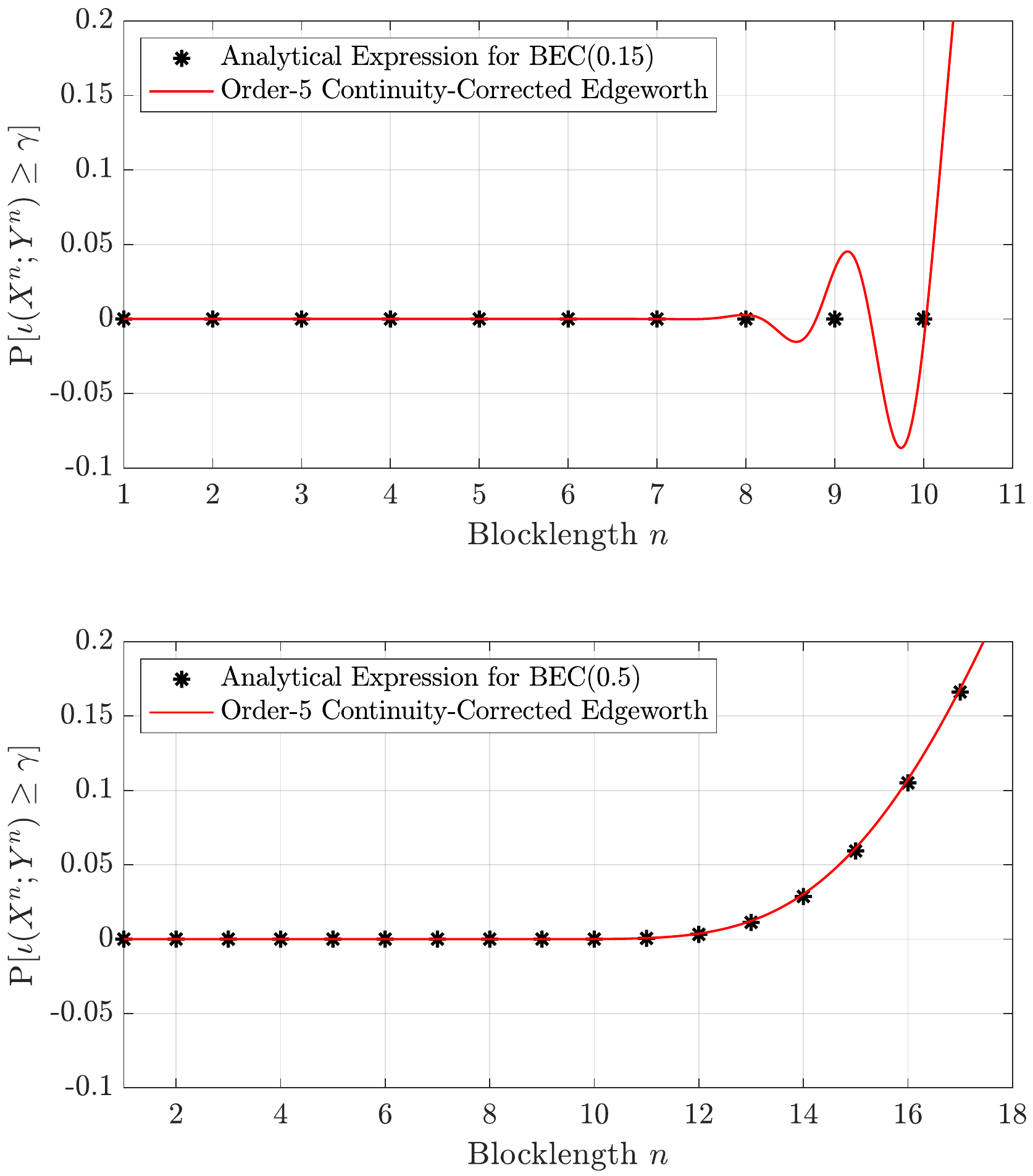}
\caption{Comparison of $\Prob[\iota(X^n; Y^n)\ge \gamma]$ between BEC$(0.15)$ and BEC$(0.5)$ with $\gamma = 10.5$. }
\label{fig: tail prob for BEC}
\end{figure}

Since $\iota(X;Y) = 1 - Z$ is a lattice random variable with span $1$, Theorem \ref{theorem: corrected Edgeworth} is readily available for approximating the tail probability $\Prob[\iota(X^n; Y^n)\ge\gamma]$. Numerical experiments show that with an order-$s$ continuity-corrected Edgeworth expansion, the oscillation issue observed in the BI-AWGN channel also persists in the BEC case. The severity of oscillation is observed to depend on the order of Edgeworth expansion $s$, the erasure probability $p$, and the choice of $\gamma$. For example, consider the continuity-corrected point $\gamma = 10.5$. Fig. \ref{fig: tail prob for BEC} shows the order-$5$ continuity-corrected Edgeworth expansions to $\Prob[\iota(X^n; Y^n)\ge\gamma]$ for BEC$(0.15)$ and for BEC$(0.5)$. We see that for BEC$(0.15)$, the approximation curve has a visible oscillation around $0$, whereas for BEC$(0.5)$, one can barely see the oscillation.

In our implementation, we choose $s = 5$ as the order for the truncated continuity-corrected Edgeworth series to approximate $\Prob[\iota(X^n; Y^n)\ge\gamma]$. For a target error probability $\epsilon = 10^{-3}$, we numerically found a sufficiently good erasure probability threshold $p^* = 0.2$, beyond which the oscillation issue becomes negligible for any continuity-corrected point $\gamma^+\ge \log\frac{1}{\epsilon} = 10$. Hence, for target error probability $\epsilon = 10^{-3}$, the $F_{\gamma}(n)$ we choose for the BEC is given by
\begin{align}
  F_{\gamma}(n) = \begin{cases}
      \sum_{c=0}^{\lfloor n - \gamma\rfloor} \binom{n}{p}p^c(1-p)^{n-c},\  n\in\N, & \text{if } p<0.2 \\
      1 - E_5(x(n); \bm{\lambda}_n),\  n\in\R_+, & \text{if } p\ge 0.2,
  \end{cases}
  \label{eq: approximation for BEC}
\end{align}
where $x(n) = \frac{n(1-p) - \gamma^+}{\sqrt{np(1-p)}}$, with $\gamma^+ = \lceil \gamma\rceil - 1/2$ being the continuity-corrected point for $\gamma$. For $p< 0.2$, we apply the discrete SDO procedure. For $p\ge 0.2$, we apply the gap-constrained SDO procedure. Both procedures are introduced in Section \ref{sec: methodology} below.


\section{An Integer Program and Two Algorithms}\label{sec: methodology}

In this section, we formulate an integer program of minimizing the upper bound on $\E[\tau]$ based on Theorem \ref{theorem: 2} and derive an error regime where Polyanskiy's stopping at zero scheme does not improve the achievability bound. Next, we provide two methods to solve this integer program: the gap-constrained SDO and the discrete SDO procedures.

By lower bounding $\Prob[\bigcup_{j=1}^i \Brace{\iota(X^{n_j}; Y^{n_j})\ge\gamma }]$ to the marginal tail probability $\Prob[\iota(X^{n_i}; Y^{n_i})\ge\gamma]$ in Theorem \ref{theorem: 2}, we write the relaxed upper bound on $\E[\tau]$ as
\begin{align}
  N(\gamma, n_1^m) &\triangleq n_m + \sum_{i=1}^{m-1}(n_i - n_{i+1})\Prob[\iota(X^{n_i}; Y^{n_i})\ge\gamma ].
\end{align}
Define the feasible region by
\begin{align}
  \F_m(\gamma, M, &\epsilon) \triangleq \{n_1^m\in\R_+^m: n_{i+1} - n_i \ge 1, \forall i\in[m-1]; \notag\\
   &\Prob[\iota(X^{n_m}; Y^{n_m})\ge\gamma]\ge 1 - \epsilon + (M-1)2^{-\gamma} \}. \label{eq: feasible region}
\end{align}
As can be seen, the feasible region \eqref{eq: feasible region} consists of real-valued decoding times $n_1^m$ such that two consecutive decoding times are separated by at least one and the target error probability is guaranteed at decoding time $n_m$.

Theorem \ref{theorem: 2} motivates the following integer program: for a given $m\in\N_+$,  $M\in\N_+$,  $\epsilon\in(0, 1)$, and $\gamma \ge \log\frac{M-1}{\epsilon}$,
\begin{align}
\begin{split}
  \min_{n_1^m}&\quad N(\gamma, n_1^m) \\
  \st&\quad n_1^m \in \F_m(\gamma, M, \epsilon) \\
    &\quad n_1^m \in \N_+^m.
\end{split}
\label{eq: integer program}
\end{align}
Let $\tilde{N}(\gamma)$ denote the minimum upper bound over $n_1^m$ after solving the integer program \eqref{eq: integer program}. Then, $\min_{\gamma}\tilde{N}(\gamma)$ yields the globally minimum upper bound $N^*(\gamma, n_1^m)$. This method is called the \emph{two-step minimization}. In this paper, we apply the two-step minimization to identify $N^*(\gamma, n_1^m)$. The key problem is to develop efficient algorithms for integer program \eqref{eq: integer program}.

In integer program \eqref{eq: integer program}, we consider $n_1\ge 1$ and integer-valued decoding times. That is, we do not allow stopping the VLSF code at $\tau = 0$. In what follows, we identify an error regime where stopping a VLSF code at $\tau = 0$ does not improve the achievability bound. 

In \cite{Polyanskiy2011}, Polyanskiy \emph{et al.} showed that the $\epsilon$-capacity $\frac{C}{1-\epsilon}$ is achievable in the non-vanishing error setting. This result is obtained by constructing a new code that stops an $(l', \N_+, M, \epsilon')$ VLSF code satisfying $\log M = Cl' + \log \epsilon' - a_0$, with $a_0 \triangleq \max_{x,y}\iota(x;y)$, at $\tau = 0$ with probability $p = \frac{\epsilon - \epsilon'}{1 - \epsilon'}$ and employs this VLSF code with probability $1 - p$, where $\epsilon'\le \epsilon$. Such a new code has probability of error $\epsilon$, average length $l = l'(1 - p)$, and message size $M^*(l, \epsilon)\ge M$. However, the following theorem shows that for sufficiently small $\epsilon$, choosing a VLSF code with $\epsilon' = \epsilon$ yields the best average length $l'$ achieved by this strategy.
\begin{theorem}\label{theorem: 7}
  Let the aforementioned notation prevail. Fix $M\in\N_{+}$. Define
    \begin{align}
      \epsilon^* \triangleq \argmin_{x\in(0, 1)}\frac{\log M + a_0 - \log x}{1 - x}. \label{eq: critical error regime}
    \end{align}
  Then, if $\epsilon \in (0, \epsilon^*]$, $\epsilon' = \epsilon$ is the minimizer that yields a minimum average length $l = l'$.
\end{theorem}

\begin{IEEEproof}
  The minimization problem we intend to solve is stated as below.
  \begin{align}
    \begin{split}
        \min_{\epsilon'}&\quad l'\Paren{1 - \frac{\epsilon - \epsilon'}{1 - \epsilon'} } \\
        \st&\quad \log M = Cl' + \log\epsilon' - a_0 \\
          &\quad \epsilon' \in (0, \epsilon].
    \end{split}
  \end{align}
This program is equivalent to the following program
  \begin{align}
    \begin{split}
      \min_{\epsilon'}&\quad \Paren{\frac{1-\epsilon}{C}}\frac{\log M + a_0 - \log \epsilon'}{1 - \epsilon'} \\
      \st&\quad \epsilon' \in (0, \epsilon].
    \end{split}
  \end{align}
Define function $f(x)\triangleq \frac{\log M + a_0 - \log x}{1 - x}$. Since $f(x)$ is convex in $(0, 1)$, there exists a unique minimizer $\epsilon^* \in (0, 1)$ for $f(x)$. Hence, if $\epsilon \le \epsilon^*$ and is fixed, the minimizer $\epsilon'  = \epsilon$, giving the minimum average length $l = l'$.
\end{IEEEproof}

Aiming to identify optimal decoding times that minimize $N(\gamma, n_1^m)$ for a given $\gamma$, we first establish a necessary condition for optimal decoding times that will aid the search for decoding times.

\begin{theorem}\label{theorem: 8}
  Let $n_0 \triangleq 0$. Denote by $S_n = \iota(X^n; Y^n)$. Fix a memoryless channel $(\X, \Y, P_{Y|X})$ and scalars $m\in\N_{+}$, $M\in\N_+$, $\epsilon\in(0, 1)$, and $\gamma \ge \log\frac{M-1}{\epsilon}$ for which the set 
  \begin{align}
  \mathsf{F}_m &\triangleq \Big\{n_1^m\in\F_m(\gamma, M, \epsilon)\cap \N_+^m: \notag\\
  &n_m = \min\{n\in\N_+: \Prob[S_n\ge\gamma]\ge 1 - \epsilon + (M-1)2^{-\gamma}\}, \notag\\
   &\forall i\in[m],\ \Prob[S_{n_i}\ge\gamma] \ge  \max_{n_{i-1}\le j < n_i}\Prob[S_j\ge\gamma] \Big\}
\end{align}
is nonempty. If $n_1^m$ are optimal decoding times for integer program \eqref{eq: integer program}, then $n_1^m\in \mathsf{F}_m$.
\end{theorem}

\begin{IEEEproof}
  Let $n_1^m\in\F_m(\gamma, M, \epsilon)\cap\N_+$ be optimal decoding times. By optimality, it satisfies $n_m = \min\{n\in\N_+: \Prob[S_n\ge\gamma]\ge 1 - \epsilon + (M-1)2^{-\gamma}\}$. Assume that for some $i\in[m-1]$, $\Prob[S_{n_i}\ge\gamma] < \max_{n_{i-1}\le j<n_i}\Prob[S_j\ge\gamma]$. In this case, we show that a better decoding time sequence in $\mathsf{F}_m$ can be constructed.

  Let $l\ge1$ be the smallest integer at which $n_l$ satisfies $\Prob[S_{n_l}\ge\gamma] < \max_{n_{l-1}\le j<n_l}\Prob[S_j\ge\gamma]$. Let $j^* = \argmax_{n_{l-1}\le j < n_l}\Prob[S_j \ge\gamma]$. There are two cases. If $n_{l-1} < j^* < n_l$, then $(n_1^{l-1}, j^*, n_{l+1}^m)$ is a better sequence than $n_1^m$. This is because
\begin{align}
  & N(\gamma, n_1^{l-1}, j^*, n_{l+1}^m) - N(\gamma, n_1^{l-1}, n_l, n_{l+1}^m) \notag\\
  &= (n_l - j^*)\Prob[S_{n_{l-1}}\ge\gamma] + (j^* - n_{l+1})\Prob[S_{j^*}\ge\gamma] \notag\\
  &\phantom{=}- (n_l - n_{l+1})\Prob[S_{n_l}\ge\gamma] \notag \\
  &< (n_l - j^*)\Prob[S_{n_{l-1}}\ge\gamma] + (j^* - n_{l+1})\Prob[S_{j^*}\ge\gamma] \notag\\
  &\phantom{=}- (n_l - n_{l+1})\Prob[S_{j^*}\ge\gamma] \notag \\
  &= (n_l - j^*)(\Prob[S_{n_{l-1}}\ge\gamma] - \Prob[S_{j^*}\ge\gamma]) \notag \\
  &\le 0.
\end{align}

If $j^* = n_{l-1}$, this implies that $\Prob[S_{n_{l-1}}\ge\gamma] > \Prob[S_{n_l}\ge\gamma]$. In this case, we employ a recursive construction of decoding time sequence as follows.
First, we replace $n_l$ with the smallest integer $\tilde{n}_l$ such that $\tilde{n}_l > n_l$ and $\Prob[S_{\tilde{n}_l}\ge\gamma] \ge \max_{n_{l-1}\le j < \tilde{n}_l}\Prob[S_j\ge\gamma]$. There are two subcases.

\textit{1)}: If $\tilde{n}_l < n_{l+1}$, then $(n_1^{l-1}, \tilde{n}_l, n_{l+1}^m)$ is a valid and better sequence than $n_1^m$. This is because
\begin{align}
  &N(\gamma, n_1^{l-1}, \tilde{n}_l, n_{l+1}^m) - N(\gamma, n_1^{l-1}, n_l, n_{l+1}^m) \notag\\
  &= (n_l - \tilde{n}_l)\Prob[S_{n_{l-1}}\ge\gamma] + (\tilde{n}_l - n_{l+1})\Prob[S_{\tilde{n}_l}\ge\gamma] \notag\\
  &\phantom{=}- (n_l - n_{l+1})\Prob[S_{n_l}\ge\gamma] \notag\\
  &< (n_l - \tilde{n}_l)\Prob[S_{n_{l-1}}\ge\gamma] + (\tilde{n}_l - n_{l+1})\Prob[S_{\tilde{n}_l}\ge\gamma] \notag\\
  &\phantom{=}- (n_l - n_{l+1})\Prob[S_{n_{l-1}}\ge\gamma] \notag\\
  &= (n_{l+1} - \tilde{n}_l)(\Prob[S_{n_{l-1}}\ge\gamma] - \Prob[S_{\tilde{n}_{l}}\ge\gamma]) \notag\\
  &\le 0.
\end{align}
The recursive construction terminates at $\tilde{n}_l$.

\textit{2)} If $\tilde{n}_l \ge n_{l+1}$, the sequence $(n_1^{l-1}, \tilde{n}_l, n_{l+1}^m)$ is invalid and the construction continues. By the construction of $\tilde{n}_l$, it follows that $\Prob[S_{\tilde{n}_l}\ge\gamma] \ge \Prob[S_{n_{l+1}}\ge\gamma]$. In this case, we replace $n_{l+1}$ with the smallest integer $\tilde{n}_{l+1}$ such that $\tilde{n}_{l+1} > \tilde{n}_l$ and $\Prob[S_{\tilde{n}_{l+1}}\ge\gamma] \ge \max_{\tilde{n}_l\le j < \tilde{n}_{l+1}}\Prob[S_j \ge\gamma] $. We check whether $\tilde{n}_{l+1} < n_{l+2}$. If true, following the similar argument as in 1), we can show $(n_1^{l-1}, \tilde{n}_l, \tilde{n}_{l+1}, n_{l+2}^m)$ is a valid and better sequence than the invalid sequence $(n_1^{l-1}, \tilde{n}_l, n_{l+1}^m)$ and the construction stops. Otherwise, we repeat the construction procedure of $\tilde{n}_{l+1}$ to construct $\tilde{n}_{l+2}$. The recursive construction stops when $\tilde{n}_t < n_{t+1}$ for some $l\le t\le m-1$. Since we are sequentially constructing better sequences, it follows that $(n_1^{l-1}, \tilde{n}_l^t, n_{t+1}^m)$ is a valid and better sequence than $n_1^m$.

For all other $n_{l'}$, $l'\ge t+1$, that satisfies $\Prob[S_{n_{l'}}\ge\gamma]< \max_{n_{l'-1}\le j<n_{l'}}\Prob[S_j\ge\gamma]$, we apply the above construction method to produce a valid and better decoding time sequence than $n_1^m$. Since $\mathsf{F}_m$ is nonempty, the construction is guaranteed to stop before $n_m$ because each time, we select the smallest integer $\tilde{n}_i$ that satisfies $\tilde{n}_i > \tilde{n}_{i-1}$ and $\Prob[S_{\tilde{n}_i}\ge\gamma]\ge \max_{\tilde{n}_{i-1}\le j < \tilde{n}_i}\Prob[S_j \ge\gamma]$. The final constructed sequence will be an element of $\mathsf{F}_m$.

To summarize, given that $\mathsf{F}_m$ is nonempty, if the optimal decoding time sequence $n_1^m$ does not belong to $\mathsf{F}_m$, then an element in $\mathsf{F}_m$ can be found that is better than $n_1^m$. Hence, $n_1^m$ must be an element of $\mathsf{F}_m$.
\end{IEEEproof}

For the BSC, Theorems \ref{theorem: 6} and \ref{theorem: 8} imply the following useful corollary.
\begin{corollary}
  Fix a BSC$(p)$, $p\in(0, 1/2)$, and scalars $m\in\N_{+}$, $M\in\N_+$, $\epsilon\in(0, 1)$ and $\gamma \ge \log\frac{M-1}{\epsilon}$. The optimal decoding times $n_1^m$ for the integer program \eqref{eq: integer program} are among the local maximizers $\{\alpha_i \}_{i=0}^\infty$.
\end{corollary}

\begin{IEEEproof}
  For brevity, let $S_n \triangleq \iota(X^n; Y^n)$. By Theorem \ref{theorem: 8}, it follows that no two optimal decoding times belong to the same interval $[\alpha_i, \alpha_{i+1})$. Otherwise, by Theorem \ref{theorem: 6}, their tail probabilities violate the condition in $\mathsf{F}_m$.

  Let us assume that there exists a sequence of decoding times $n_1^m \in \F_m(\gamma, M, \epsilon)$ for BSC$(p)$ satisfying $n_i \in [\alpha_{b_i}, \alpha_{b_i+1})$, $i\in[m]$, where $b_1 < b_2 < \cdots < b_m$. If $n_i > \alpha_{b_i}$, then by Theorem \ref{theorem: 6}, $\Prob[S_{\alpha_{b_i}}\ge\gamma] > \Prob[S_{n_i}\ge\gamma]$. This violates the necessary condition in Theorem \ref{theorem: 8}. Hence, if $n_1^m$ are optimal, we must have $n_i = \alpha_{b_i}$.
\end{IEEEproof}

Finally, we present two search algorithms to numerically solve integer program \eqref{eq: integer program}: the gap-constrained SDO procedure and the discrete SDO procedure. The first algorithm relies on a monotone, differentiable function $F_{\gamma}(n)$, $n\ge 0$, to approximate $\Prob[\iota(X^n; Y^n)\ge\gamma]$. In contrast, the second algorithm only relies on a good estimate of $\Prob[\iota(X^n; Y^n)\ge\gamma]$, $n\in\N_+$, at a cost of increased search complexity.

\subsection{The Gap-Constrained SDO Procedure}

To facilitate a program that is computationally tractable, we consider the relaxed program of \eqref{eq: integer program} by allowing $n_1^m \in\R_+^m$: For a given $m\in\N_+$, $M\in\N_+$, $\epsilon \in (0, 1)$, and $\gamma \ge \log\frac{M - 1}{\epsilon}$,
\begin{align}
  \begin{split}
    \min_{n_1^m}&\quad N(\gamma, n_1^m) \\
    \st&\quad n_1^m\in \F_m(\gamma, M, \epsilon).
  \end{split}
\label{eq: relaxed program}
\end{align}
In the relaxed program \eqref{eq: relaxed program}, the tail probability $\Prob[\iota(X^n; Y^n)\ge\gamma]$ is approximated by a monotonically increasing and differentiable function $F_{\gamma}(n)$ satisfying $F_{\gamma}(0) = 0$ and $F_{\gamma}(\infty) = 1$. Define
\begin{align}
  f_{\gamma}(n) \triangleq \frac{\diff F_{\gamma}(n)}{\diff n}.
\end{align}
The next theorem gives the analytical solution to the relaxed program \eqref{eq: relaxed program}.

\begin{theorem}\label{theorem: GC-SDO}
  Fix a memoryless channel $(\X, \Y, P_{Y|X})$ for which $\iota(X;Y)$ is continuous and $\Prob[\iota(X^n; Y^n)\ge\gamma]$ is strictly increasing. For a given $m\in\N_+$, $M\in\N_+$, $\epsilon\in(0, 1)$, and $\gamma \ge \log\frac{M-1}{\epsilon}$, let $\bar{n} = F_{\gamma}^{-1}\Paren{1 - \epsilon + (M-1)2^{-\gamma} }$. If $\bar{n} > m-1$ and $\sum_{i=1}^{m-1}f_{\gamma}(x-i)<1$ for all $x\ge\bar{n}$, then the optimal real-valued decoding times $n_1^*, n_2^*, \dots, n_m^*$ for the relaxed program satisfy
  \begin{align}
    &n_m^* = \bar{n}, \label{eq: GC_SDO_eq1} \\
    & n_{i+1}^* = n_i^* + \max\Brace{1,\frac{F_{\gamma}(n_i^*) - F_{\gamma}(n_{i-1}^*) - \lambda_{i-1} }{f_{\gamma}(n_i^*)}}, \label{eq: GC_SDO_eq2} \\
    & \lambda_i = \max\{\lambda_{i-1}+f_{\gamma}(n_i^*) - F_{\gamma}(n_i^*) + F_{\gamma}(n_{i-1}^*), 0 \}, \label{eq: GC_SDO_eq3}
  \end{align}
  where $i\in[m-1]$, $\lambda_0\triangleq 0$, and $n_0^*\triangleq 0$.
\end{theorem}

\begin{remark}
    For BI-AWGN channels, $F_{\gamma}(n)$ is typically convex for small $n$ and becomes concave for large $n$. For such $F_{\gamma}(n)$, let $x_I$ be the single inflection point of $F_{\gamma}(n)$. It suffices to examine the condition $\sum_{i=1}^{m-1}f_{\gamma}(x-i)<1$ for $x\in [x_I, x_I + m - 1]$.
\end{remark}

\begin{IEEEproof}
    For brevity, denote by $\bm{n} \triangleq (n_1, n_2, \dots, n_m)$ the vector of decoding times and by $\bm{n}^*\triangleq (n_1^*, n_2^*, \dots, n_m^*)$ the vector of optimal decoding times. By introducing the Lagrangian multipliers $\nu$, $\lambda_1^{m-1}$, the Lagrangian of the relaxed program is given by
\begin{align}
    \LL(\bm{n}, \nu, &\lambda_1^{m-1}) = n_m + \nu(1 - F_{\gamma}(n_m) - \epsilon + (M-1)2^{-\gamma})  \notag\\
    & + \sum_{i=1}^{m-1}(n_{i} - n_{i+1})F_{\gamma}(n_i) + \sum_{i=1}^{m-1}\lambda_i(n_i - n_{i+1} + 1). \notag
\end{align}
The optimal decoding times $\bm{n}^* = (n_1^*, n_2^*,\dots, n_m^*)$ must satisfy the Karush-Kuhn-Tucker (KKT) conditions \cite[Sec. 5.5.3]{Boyd}: $\nu\ge0$, $\lambda_i\ge0$, $i\in[m-1]$,
\begin{align}
    &\frac{\partial \LL}{\partial n_1}\Big|_{\bm{n}=\bm{n}^*} = F_{\gamma}(n^*_1) - (n^*_2-n^*_1)f_{\gamma}(n_1^*) + \lambda_1 = 0\\
    &\frac{\partial \LL}{\partial n_i}\Big|_{\bm{n}=\bm{n}^*} = F_{\gamma}(n_i^*) - F_{\gamma}(n_{i-1}^*) - (n_{i+1}^* - n_i^*)f_{\gamma}(n_i^*) \notag\\
    &\phantom{\frac{\partial \LL}{\partial n_k}\Big|_{\bm{n}=\bm{n}^*} =}+ \lambda_i - \lambda_{i-1} = 0,\quad 2\le i\le m-1, \label{eq: th8_1}\\
    &\frac{\partial \LL}{\partial n_m}\Big|_{\bm{n}=\bm{n}^*} = 1 - F_{\gamma}(n_{m-1}^*) - \nu f_{\gamma}(n_m^*) - \lambda_{m-1} = 0, \label{eq: th8_2}\\
    &\nu(1 - F_{\gamma}(n_m^*) -\epsilon + (M-1)2^{-\gamma}) = 0, \label{eq: th8_3} \\
  &\lambda_i(n_i^* - n_{i+1}^* + 1) = 0,\quad i\in[m-1]. \label{eq: th8_4}
\end{align}

We analyze \eqref{eq: th8_4}. There are two cases. If $\lambda_i > 0$, then $n^*_{i+1} = n_i^* + 1$. By \eqref{eq: th8_1}, we obtain
\begin{align}
  \lambda_i &= \lambda_{i-1}+f_{\gamma}(n_i^*) - F_{\gamma}(n_i^*) + F_{\gamma}(n_{i-1}^*),\ 2\le i\le m-1\\
  \lambda_1 &= f_{\gamma}(n_1^*) - F_{\gamma}(n_1^*)
\end{align}
If $n_{i+1}^* > n_i^* + 1$, then $\lambda_i = 0$. By \eqref{eq: th8_1}, we obtain
\begin{align}
    n_{i+1}^* = n_i^* + \frac{F_{\gamma}(n_i^*) - F_{\gamma}(n_{i-1}^*) - \lambda_{i-1} }{f_{\gamma}(n_i^*)}.
\end{align}
Rewriting the above two cases in a compact form yields \eqref{eq: GC_SDO_eq2} and \eqref{eq: GC_SDO_eq3}.

Next, we prove \eqref{eq: GC_SDO_eq1}. Let $n_m^* \ge \bar{n}$ be fixed which guarantees $F_{\gamma}(n_m^*)\ge 1 - \epsilon + (M-1)2^{-\gamma}$. We wish to maximize $\lambda_{m-1}$ and show that $\max\{\lambda_{m-1}\}<1 - F_{\gamma}(n_{m-1}^*)$. Assume that $\lambda_{m-1}>0$. Then by the above analysis,
\begin{align}
  &n^*_{m-1} = n^*_{m} - 1 \\
  &\lambda_{m-1} = (\lambda_{m-2}+ F_{\gamma}(n_{m-2}^*)) +f_{\gamma}(n_{m-1}^*) - F_{\gamma}(n_{m-1}^*) .
\end{align}
It follows that $\lambda_{m-1}$ is maximized if and only if $\lambda_{m-2} + F_{\gamma}(n_{m-2}^*)$ is maximized. Since $F_{\gamma}(n)$ is strictly increasing and $n^*_{m-2}\le n^*_{m-1}-1$, the maximum of $\lambda_{m-2} + F_{\gamma}(n_{m-2}^*)$ is achieved by $\lambda_{m-2} > 0$ and $n^*_{m-2} = n^*_{m-1} - 1$. Hence,
\begin{align}
  &n^*_{m-2} = n^*_{m-1} - 1 \\
  &\lambda_{m-2} = (\lambda_{m-3}+ F_{\gamma}(n_{m-3}^*)) +f_{\gamma}(n_{m-2}^*) - F_{\gamma}(n_{m-2}^*) .
\end{align}
Repeating the above analysis to $\lambda_{i}+F_{\gamma}(n_i)$ for all $2\le i\le m-3$, we get
\begin{align}
  &n^*_{i} = n^*_{i+1} - 1 \\
  &\lambda_{i} = (\lambda_{i-1}+ F_{\gamma}(n_{i-1}^*)) + f_{\gamma}(n_{i}^*) - F_{\gamma}(n_{i}^*).
\end{align}
and $\lambda_1 = f_{\gamma}(n_1^*) - F_{\gamma}(n^*_1) > 0$.
Substituting $\lambda_{i}$ into the expression of $\lambda_{i+1}$ recursively, we obtain the maximum value of $\lambda_{m-1}$ given by
\begin{align}
  \lambda_{m-1} &= \sum_{j=1}^{m-1}f_{\gamma}(n^*_{m-j}) - F_{\gamma}(n^*_{m-1}) \notag\\
  &= \sum_{j=1}^{m-1}f_{\gamma}(n^*_m-j) - F_{\gamma}(n^*_{m}-1).
\end{align}
Plugging this into \eqref{eq: th8_2} and solving for $\nu$, we obtain that when $\lambda_{m-1}$ is maximized,
\begin{align}
    \nu = \frac{1 - \sum_{i=1}^{m-1}f_{\gamma}(n^*_m - i)}{f_{\gamma}(n^*_m)}. \label{eq: nu formula}
\end{align}
Invoking the conditions that $\bar{n}>m-1$ and $\sum_{i=1}^{m-1}f_{\gamma}(x - i)<1$ for all $x\ge \bar{n}$, it follows that $\nu > 0$. Thus, by \eqref{eq: th8_3}, we conclude that $n_m^* = \bar{n}$, which completes the proof of \eqref{eq: GC_SDO_eq1}.
\end{IEEEproof}

The procedures \eqref{eq: GC_SDO_eq1} -- \eqref{eq: GC_SDO_eq3} are called the \emph{gap-constrained SDO procedure}. The name indicates that the solution ensures two consecutive decoding times are separated by at least one. In contrast, the \emph{unconstrained} SDO procedure considered in previous works \cite{Vakilinia2016,Wang2017,Wesel2018,Wong2017,Heidarzadeh2018,Heidarzadeh2019} does not consider the gap constraint and admits a simple recursion
\begin{align}
  n_{i+1}^* = n_i^* + \frac{F_{\gamma}(n_i^*) - F_{\gamma}(n_{i-1}^*) }{f_{\gamma}(n_i^*) }, \quad i\in[m-1] \label{eq: unconstrained SDO}
\end{align}
with $n_m^*$ determined by \eqref{eq: GC_SDO_eq1}. 

\begin{figure}[t]
\centering
\includegraphics[width=0.48\textwidth]{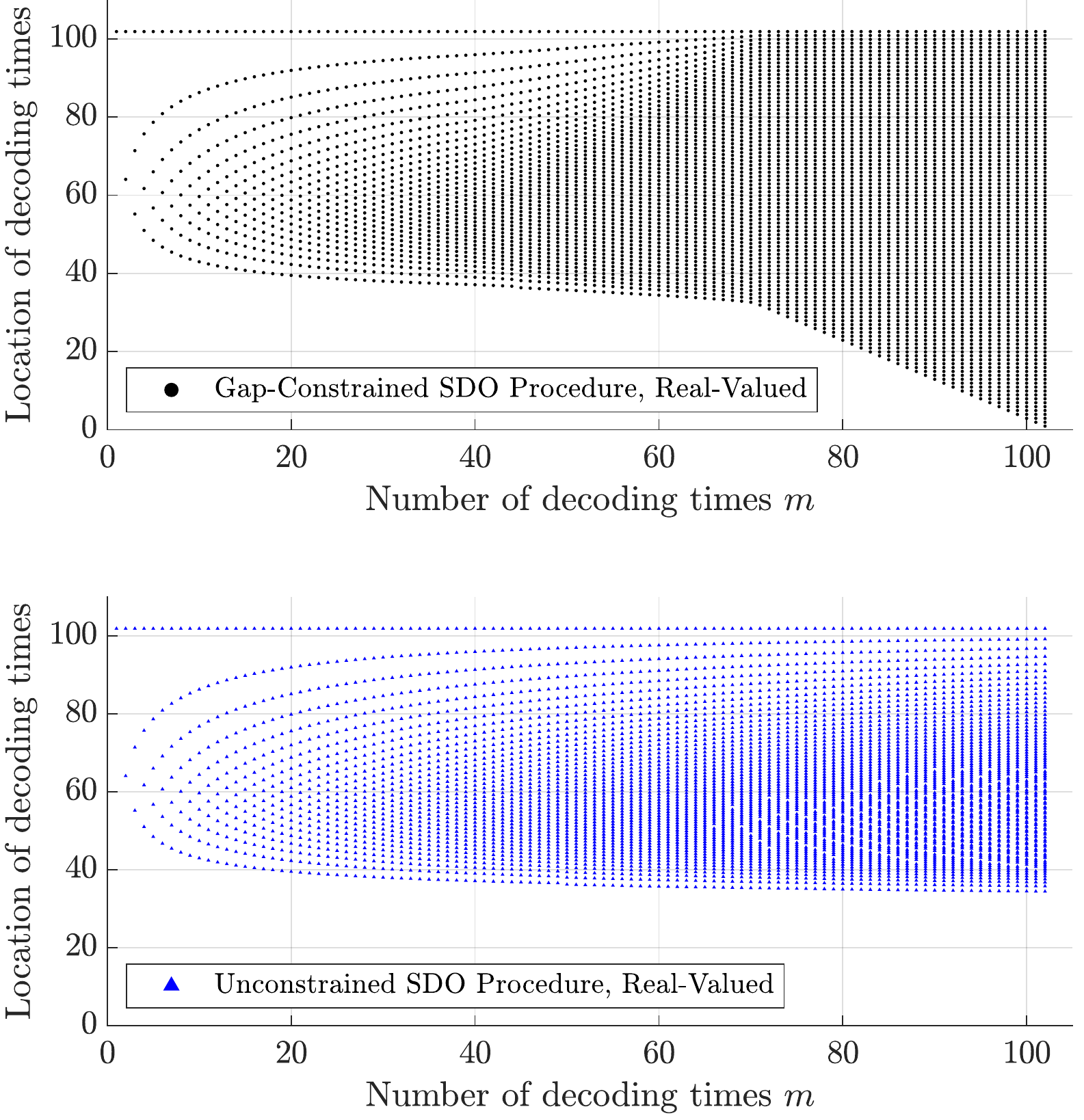}
\caption{Comparison of the optimal real-valued decoding times between the gap-constrained and unconstrained SDO procedures for the BI-AWGN channel at $0.2$ dB. In this example, we choose $\epsilon = 10^{-2}$, $(M - 1)2^{-\gamma} = \epsilon/2$, where $M = 2^{20}$. This produces $\gamma = 27.64$ and $n_m^* = 101.91$ using \eqref{eq: GC_SDO_eq1}. We consider number of decoding times $m$ ranging from $1$ to $\lceil n_m^* \rceil = 102$. }
\label{fig: decoding time evolution}
\end{figure}

To illustrate the distinction between the gap-constrained and unconstrained SDO procedures, Fig. \ref{fig: decoding time evolution} shows how the optimal real-valued decoding times $n_1^m$ evolves as $m$ increases using these two algorithms for the BI-AWGN channel at $0.2$ dB, $M = 2^{20}$, and $\epsilon = 10^{-2}$. Together, these parameters uniquely determine $n_m^* = 101.91$ via \eqref{eq: GC_SDO_eq1}. For $m \le 20$, the gap-constrained SDO procedure behaves indistinguishably from the unconstrained SDO procedure, since the SDO solution naturally has a minimum gap larger than one. For large values of $m$, the unconstrained SDO procedure avoids early decoding times and instead adds later decoding times so densely that their separation is less than one. In contrast, the gap-constrained SDO procedure is forced to add early decoding times when all existing gaps become one.

We remark that the form of the gap-constrained SDO procedure naturally calls for a bisection search to identify $n_1^*$ that subsequently determines $n_2^*, \dots, n_{m-1}^*$. When evaluating at small values of $n$, both $F_{\gamma}(n)$ and $f_{\gamma}(n)$ will become infinitesimally small. In this case, a direct numerical computation using \eqref{eq: GC_SDO_eq2} and \eqref{eq: GC_SDO_eq3} may cause a precision issue. Fortunately, the gap-constrained SDO procedures also admit a ratio form.  Define $\lambda_k^{(r)} \triangleq \lambda_k / f_{\gamma}(n_k^*) $. Then, \eqref{eq: GC_SDO_eq2} and \eqref{eq: GC_SDO_eq3} can be equivalently written as
\begin{align}
    & n_{i+1}^* = n_i^* \notag\\
    &\phantom{n}+ \max\Bigg\{1, \frac{F_{\gamma}(n_i^*)}{f_{\gamma}(n_i^*)} - \frac{F_{\gamma}(n_{i-1}^*)}{f_{\gamma}(n_i^*)} - \lambda_{i-1}^{(r)}\frac{f_{\gamma}(n_{i-1}^*)}{f_{\gamma}(n_i^*)} \Bigg\},  \\
    & \lambda_i^{(r)} \notag\\
    &= \max\Bigg\{\lambda_{i-1}^{(r)}\frac{f_{\gamma}(n_{i-1}^*)}{f_{\gamma}(n_i^*)} + 1 - \frac{F_{\gamma}(n_i^*)}{f_{\gamma}(n_i^*)} + \frac{F_{\gamma}(n_{i-1}^*)}{f_{\gamma}(n_i^*)}, 0\Bigg\}.  
\end{align}
The purpose of using $F_{\gamma}(\tilde{n})/f_{\gamma}(n)$, $f_{\gamma}(\tilde{n})/f_{\gamma}(n)$, and $\lambda_k^{(r)}$ is that they have a closed-form expression that cancels out the common infinitesimal factor in both the numerator and denominator. In our implementation for the BI-AWGN channel, we applied the ratio form of the gap-constrained SDO procedure.

\subsection{The Discrete SDO Procedure}

The gap-constrained SDO procedure in Theorem \ref{theorem: GC-SDO} hinges on the existence of a monotonically increasing and differentiable function $F_{\gamma}(n)$ to approximate $\Prob[\iota(X^n; Y^n)\ge\gamma]$. In general, however, such a function may not exist. For example, for the BSC$(p)$, $p\in(0, 1/2)$, the tail probability $\Prob[\iota(X^n; Y^n)\ge\gamma]$ as a function of $n$ cannot be approximated by a monotone and differentiable function, as seen in Fig. \ref{fig: tail prob for BSC}.

As a general solution to the integer program \eqref{eq: integer program}, we develop the \emph{discrete SDO procedure} that only relies on a good estimate of $\Prob[\iota(X^n; Y^n)\ge\gamma]$ at $n\in\N_+$. 
\begin{theorem} \label{theorem: discrete SDO}
  Fix a memoryless channel $(\X, \Y, P_{Y|X})$ and scalars $m\in\N_+$, $M\in\N_+$, $\epsilon\in(0, 1)$, and $\gamma \ge \log\frac{M-1}{\epsilon}$. Define $S_n\triangleq \iota(X^n; Y^n)$. The optimal integer-valued decoding times $n_1^*, n_2^*, \dots, n_m^*$ for the integer program \eqref{eq: integer program} satisfy
    \begin{align}
      &n_1^* + \max(1,g_{-}^{(1)}(n_1^*)) \le n_2^* \le n_1^* + g_{+}^{(1)}(n_1^*), \label{eq: discrete_SDO_eq1} \\
      &n_{i}^* + \max(1, g_{-}^{(i)}(n_{i}^*, n_{i-1}^*)) \le n_{i+1}^*  \le n_{i}^* + g_{+}^{(i)}(n_{i}^*, n_{i-1}^*), \notag\\ &\text{for } i\in\{2,3,\dots, m-1\}, \label{eq: discrete_SDO_eq3}
    \end{align}
  where $n_m^*$ is the smallest integer $n_m$ at which $\Prob[S_{n_m}\ge\gamma]\ge 1 - \epsilon + (M - 1)2^{-\gamma}$. For $n_1$, the $g$ functions associated with $n_1$ are defined by
  \begin{align}
    g_{-}^{(1)}(n_1) &\triangleq \max_{\substack{n\in[1, n_m^*-m+1] \\ \Prob[S_n\ge\gamma] < \Prob[S_{n_1}\ge\gamma] } } \frac{\Prob[S_{n}\ge\gamma](n_1 - n)}{\Prob[S_{n_1}\ge\gamma] - \Prob[S_{n}\ge\gamma]}, \\
    g_{+}^{(1)}(n_1) &\triangleq \min_{\substack{n\in[1, n_m^*-m+1] \\ \Prob[S_n\ge\gamma] > \Prob[S_{n_1}\ge\gamma] } } \frac{\Prob[S_{n}\ge\gamma](n_1 - n) }{\Prob[S_{n_1}\ge\gamma] - \Prob[S_{n}\ge\gamma]}.
  \end{align}
  For $2\le i\le m-1$, the $g$ functions associated with the pair $(n_i, n_{i-1})$ are defined by
  \begin{align}
    &g_{-}^{(i)}(n_i, n_{i-1}) \notag\\
    &\triangleq \max_{\substack{n\in[n_{i-1}+1, n_m^*-m+i] \\ \Prob[S_n\ge\gamma] < \Prob[S_{n_i}\ge\gamma] }} \frac{\Prob[S_{n}\ge\gamma] - \Prob[S_{n_{i-1}}\ge\gamma]}{\Prob[S_{n_i}\ge\gamma] - \Prob[S_{n}\ge\gamma]}(n_i - n), \\
  &g_{+}^{(i)}(n_i, n_{i-1}) \notag\\
  &\triangleq \min_{\substack{n\in[n_{i-1}+1, n_m^*-m+i] \\ \Prob[S_n\ge\gamma] > \Prob[S_{n_i}\ge\gamma] }} \frac{\Prob[S_{n}\ge\gamma] - \Prob[S_{n_{i-1}}\ge\gamma] }{\Prob[S_{n_i}\ge\gamma] - \Prob[S_{n}\ge\gamma]}(n_i - n).
  \end{align}
  For $g_{-}$ functions defined above, if the maximizer is empty, $g_{-}^{(i)}(\cdot) = -\infty$, $i\in[m-1]$. For $g_{+}$ functions defined above, if the minimizer is empty, $g_{+}^{(i)}(\cdot) = \infty$, $i\in[m-1]$.
\end{theorem}

\begin{IEEEproof}
  Since the $m$th decoding time is used to meet the target error probability, it follows that the optimal $n_m^*$ corresponds to the smallest integer $n_m$ at which $\Prob[S_{n_m}\ge\gamma]\ge 1 - \epsilon + (M - 1)2^{-\gamma}$.

  Assume that $n_1^*, n_2^*,\dots, n_m^*$ are optimal decoding times. This means that for any other $n_1$, we have
  \begin{align}
    N(\gamma, n_1^*, n_2^*,\dots,n_m^*) \le N(\gamma, n_1, n_2^*,\dots,n_m^*). \label{eq: th9_eq1}
  \end{align}
  Inequality \eqref{eq: th9_eq1} is equivalent to the following
  \begin{align}
    (n_1^* - n_2^*)\Prob[S_{n_1^*}\ge\gamma] - (n_1 - n_2^*)\Prob[S_{n_1}\ge\gamma] \le 0. \label{eq: th9_eq2}
  \end{align}
We distinguish two cases. If $\Prob[S_{n_1}\ge\gamma] < \Prob[S_{n_1^*}\ge\gamma]$, \eqref{eq: th9_eq2} is equivalent to
  \begin{align}
    n_2^* \ge n_1^* + \frac{\Prob[S_{n_1}\ge\gamma](n_1^* - n_1)}{\Prob[S_{n_1^*}\ge\gamma] - \Prob[S_{n_1}\ge\gamma]}. \label{eq: th9_eq3}
  \end{align}
If $\Prob[S_{n_1^*}\ge\gamma] < \Prob[S_{n_1}\ge\gamma]$, \eqref{eq: th9_eq2} is equivalent to
  \begin{align}
    n_2^* \le n_1^* + \frac{\Prob[S_{n_1}\ge\gamma](n_1^* - n_1)}{\Prob[S_{n_1^*}\ge\gamma] - \Prob[S_{n_1}\ge\gamma]}. \label{eq: th9_eq4}
  \end{align}
Note that $n_2^*$ should satisfy  \eqref{eq: th9_eq3} for all $n_1\in[1, n_m^* - m + 1]$ with $\Prob[S_{n_1}\ge\gamma] < \Prob[S_{n_1^*}\ge\gamma]$. Using the $g_{-}^{(1)}$  function defined earlier and noting that $n_2^*\ge n_1^* + 1$, \eqref{eq: th9_eq3} can be compactly written as
  \begin{align}
    n_2^* \ge n_1^* + \max(1, g_{-}^{(1)}(n_1^*)).
  \end{align}
In a similar fashion, \eqref{eq: th9_eq4} can be written as
  \begin{align}
    n_2^* \le n_1^* + g_{+}^{(1)}(n_1^*).
  \end{align}
The necessary conditions for $n_3^*,\dots, n_m^*$ can be derived analogously.
\end{IEEEproof}

Inequalities \eqref{eq: discrete_SDO_eq1} to \eqref{eq: discrete_SDO_eq3} are called the \emph{discrete SDO procedure} due to their resemblance to the unconstrained SDO update rule \eqref{eq: unconstrained SDO} and the discrete nature. Note that the discrete SDO procedure automatically meets the desired gap constraint. Furthermore, the discrete SDO procedure is in fact a depth-first search which may produce a collection of decoding times $n_1^m$ that satisfy inequalities \eqref{eq: discrete_SDO_eq1} to \eqref{eq: discrete_SDO_eq3}. In this case, the optimal decoding times are among these finalists that yield the minimum upper bound $\tilde{N}(\gamma)$ for a fixed $\gamma$.

In general, the complexity of the discrete SDO procedure is much higher than that of the gap-constrained SDO procedure when $\Prob[\iota(X^n; Y^n)\ge\gamma]$ is an increasing function of $n$. Nevertheless, the discrete SDO procedure could become much more efficient if the search for optimal decoding times is restricted to a sparse set that meets Theorem \ref{theorem: 8}, for instance, the set of local maximizers $\Brace{\alpha_i}_{i=1}^\infty$ in the BSC case.

\section{Numerical Evaluations}\label{sec: numerical evaluations}

This section numerically evaluate the achievability bound of VLSF codes with finite decoding times  for three important channels: the BI-AWGN channel, the BSC, and the BEC. For each channel, we apply distinct computational methods, but we always use the two-step minimization in Sec. \ref{sec: methodology}  to obtain the globally minimum upper bound $N^*(\gamma, n_1^m)$. We then use $N^*(\gamma, n_1^m)$  to obtain  the achievability bound on rate given by $\frac{\log M}{N^*(\gamma, n_1^m)}$, where $M$ is the message size.

\begin{figure}[t]
\centering
\includegraphics[width=0.48\textwidth]{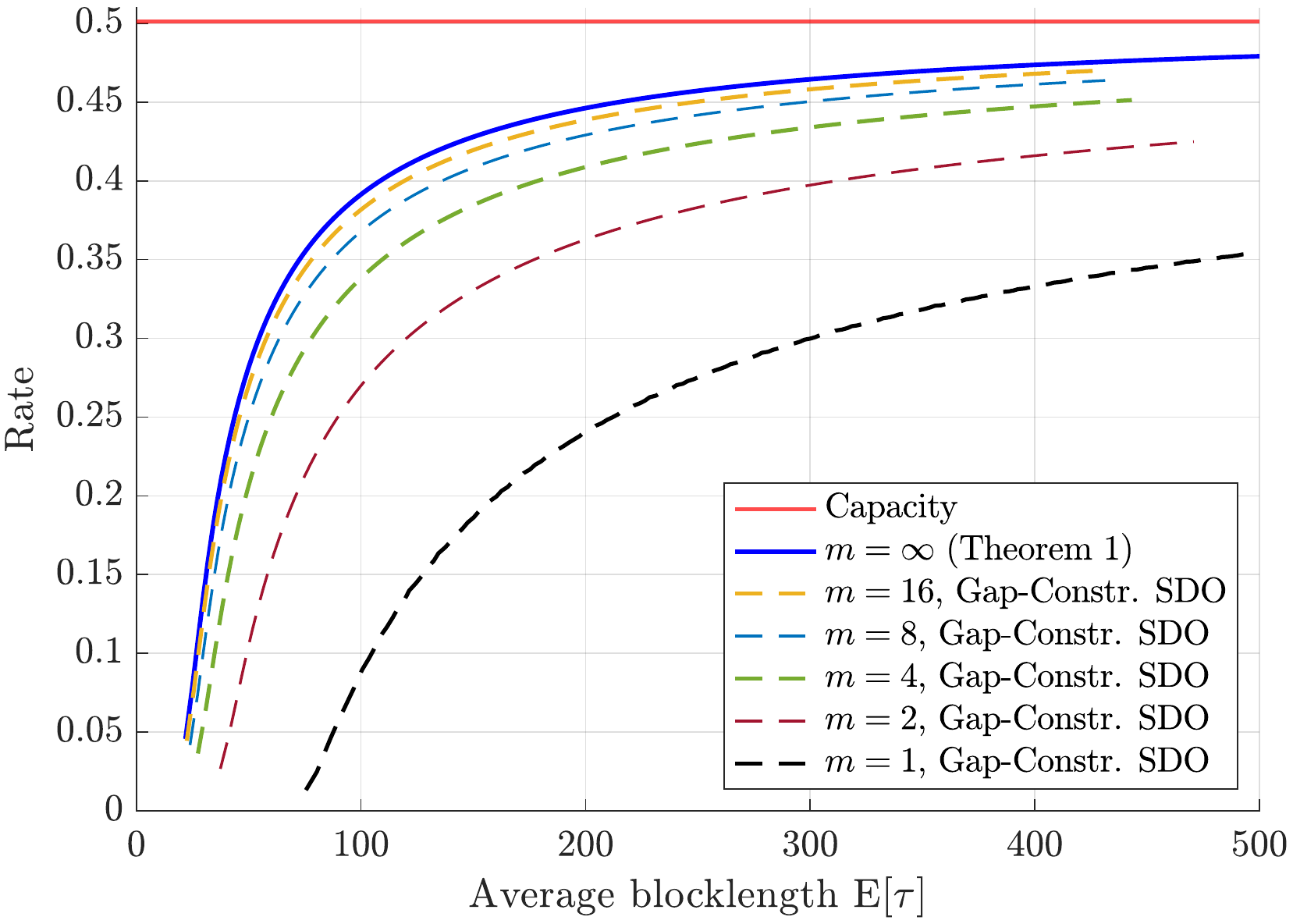}
\caption{Rate vs. average blocklength $\E[\tau]$ for BI-AWGN channel at $0.2$ dB and $\epsilon = 10^{-3}$. In this example, $k$ ranges from $1$ to $200$.}
\label{fig: rate for BI-AWGN}
\end{figure}

\begin{figure}[t]
\centering
\includegraphics[width=0.48\textwidth]{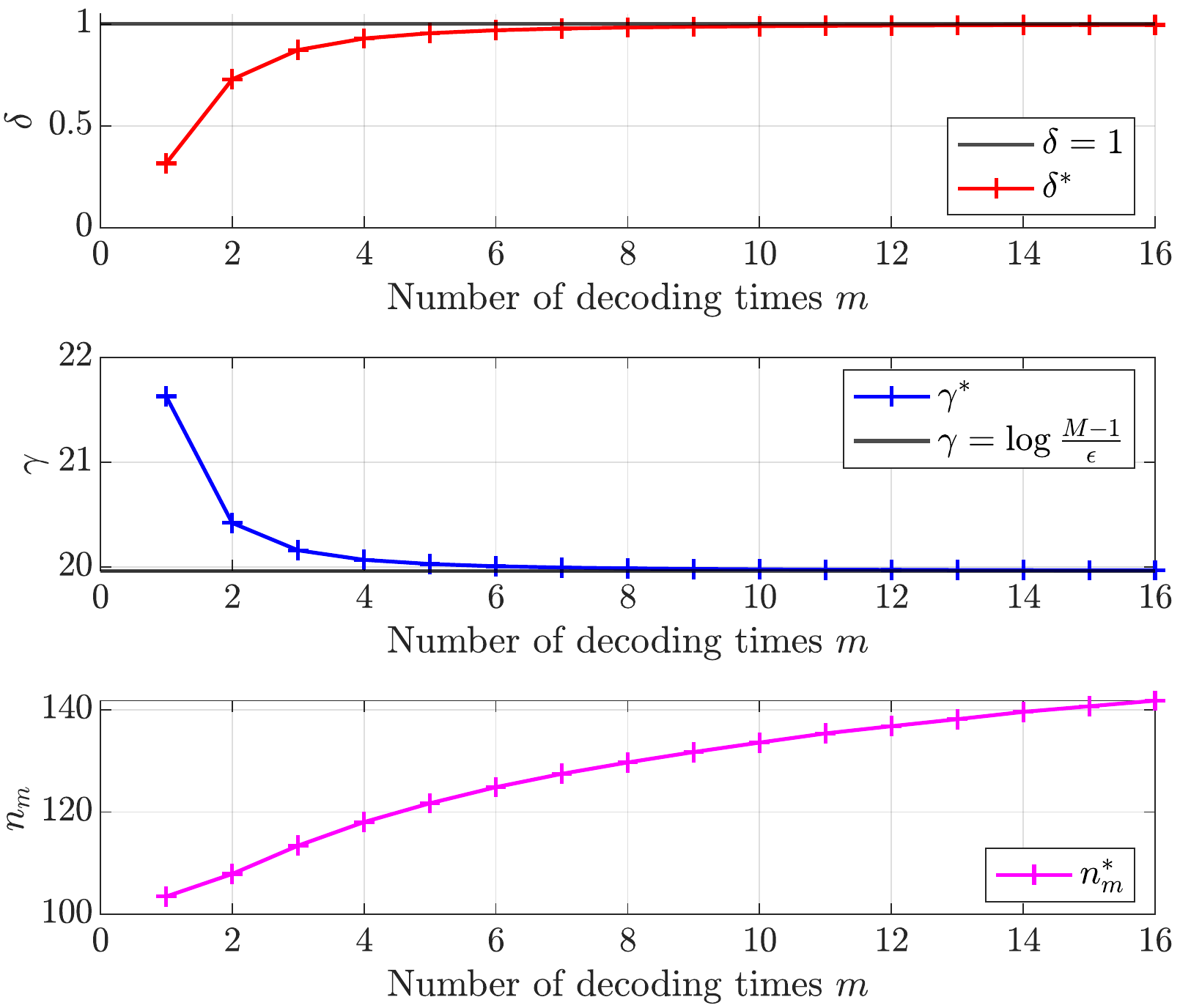}
\caption{Globally optimal $\delta^*$, $\gamma^*$, and $n_m^*$ as a function of the number of decoding times $m$ for BI-AWGN channel at $0.2$ dB, $k = 10$, and $\epsilon = 10^{-3}$. }
\label{fig: params for BI-AWGN}
\end{figure}

We consider the error regime in which Polyanskiy's stopping-at-zero scheme does not improve the achievability bound, which is identified by Theorem \ref{theorem: 7}. Denote by $k \triangleq \log M$ the information length. Note that the three binary-input channels have maximum information density $a_0 \in\{1, \log2(1-p)\}$.  Thus, for  $k\le 1000$, numerical evaluation of \eqref{eq: critical error regime} shows that $\epsilon \le 1.4\times 10^{-3}$ is the error regime in which Polyanskiy's stopping-at-zero scheme does not improve the achievability bound for any of the three binary-input channels. Throughout this section, we consider a fixed target error probability $\epsilon = 10^{-3}$ which falls into the above error regime for $k\le 1000$. In this section, we use \eqref{eq: relaxation on blocklength} and \eqref{eq: relaxation on error prob} to numerically evaluate Polyanskiy's achievability bound on VLSF codes in Theorem \ref{theorem: 1} for the three binary-input channels.

\subsection{BI-AWGN Channel}\label{subsec: sim for BI-AWGN}

We consider the BI-AWGN channel at SNR $0.2$ dB, so that the capacity $C = 0.5$ bits/channel use. The approximation function $F_{\gamma}(n)$ that we use to approximate $\Prob[\iota(X^n; Y^n)\ge\gamma]$ is given by \eqref{eq: approx function for BI-AWGN}, namely, a combination between the order-$5$ Edgeworth expansion and the order-$3$ Petrov expansion. Although the derivative at $n^*$ in \eqref{eq: approx function for BI-AWGN} is unspecified, one can define its derivative as its left or right derivative and this does not affect the SDO performance.  We apply the gap-constrained SDO procedure to solve the relaxed program \eqref{eq: relaxed program}.

For $\epsilon = 10^{-3}$ and the BI-AWGN channel at $0.2$ dB, Fig. \ref{fig: rate for BI-AWGN} shows achievability bounds estimated by the gap-constrained SDO procedure and two-step minimization for $m = 1, 2, 4, 8, 16$. When $m$ is small, a slight increase in $m$ dramatically improves the achievability bound of the VLSF code. However, this improvement is diminishing as $m$ gets large enough. We see that Polyanskiy's achievability bound can be closely  approached with $m = 16$ for a wide range of average blocklength (or $k$).

One may wonder the following problem: for a fixed $k$, how do the optimal $\gamma^*$ and $n_m^*$ evolve as $m$ increases? By introducing a new parameter $\delta\in(0, 1)$ to \eqref{eq: GC_SDO_eq1}, we assign $\delta\epsilon$ error probability to the term $(M-1)2^{-\gamma}$ so that 
\begin{align}
  \gamma(\delta) &= \log\frac{M-1}{\delta\epsilon}, \label{eq: gamma_delta} \\
  n_m(\delta) &= F_{\gamma(\delta)}^{-1}(1 - \epsilon + \delta\epsilon). \label{eq: n_m_delta}
\end{align} 
Fig. \ref{fig: params for BI-AWGN} shows how the optimal $\delta^*$ evolves as $m$ increases for $k = 10$ and $\epsilon = 10^{-3}$ for the BI-AWGN channel at $0.2$ dB. Using \eqref{eq: gamma_delta} and \eqref{eq: n_m_delta}, $\gamma^*$ and $n_m^*$ are uniquely determined. We observe that for $m = 1$, $\delta^* < 1/2$. As $m$ increases, $\delta^*$ quickly approaches one. This drives $\gamma$ to approach $\log\frac{M-1}{\epsilon}$, and $n_m$ to $\infty$. This trend matches Polyanskiy's setup for an $(l,\N, M, \epsilon)$ VLSF code.


\begin{figure}[t]
\centering
\includegraphics[width=0.48\textwidth]{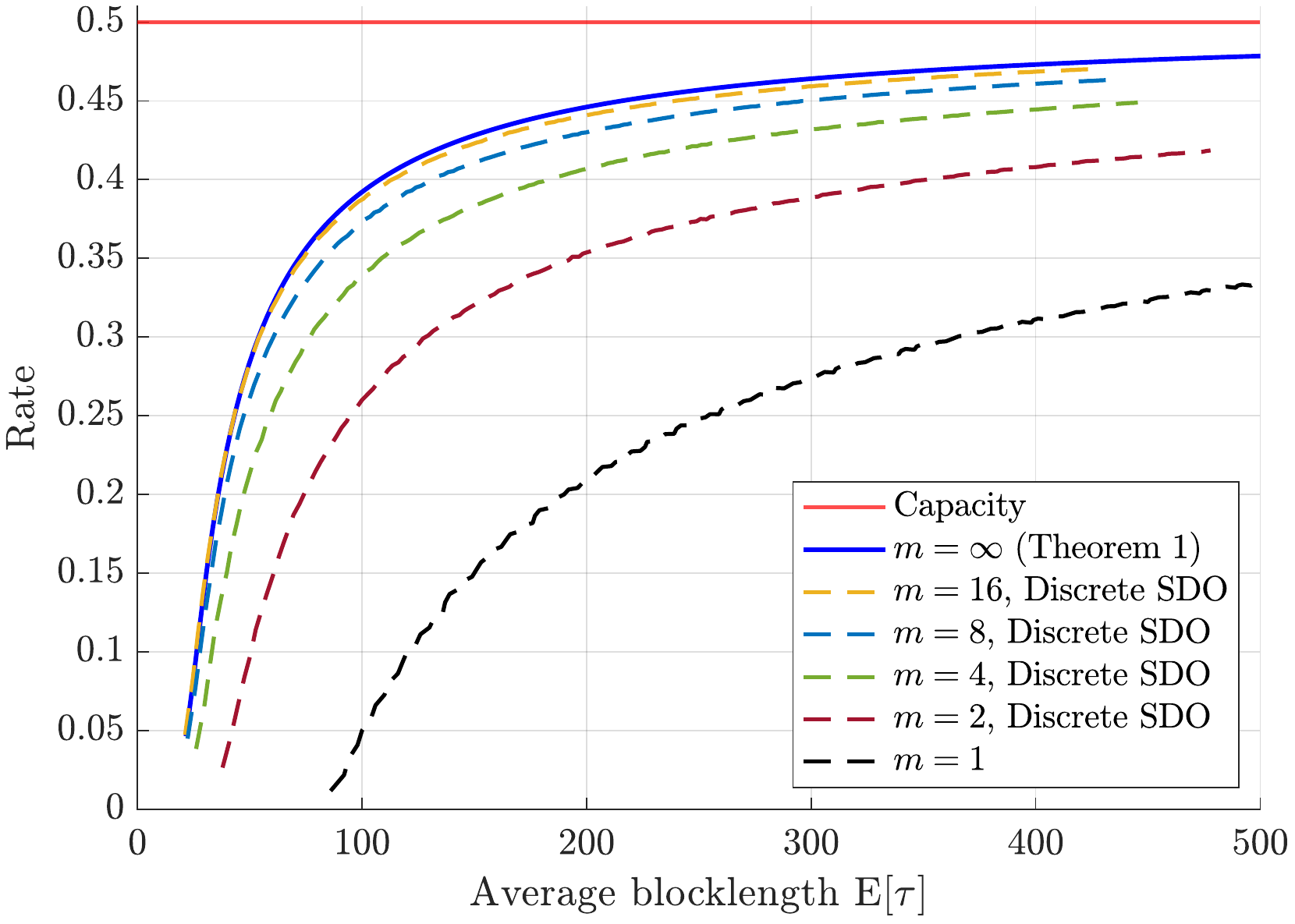}
\caption{Rate vs. average blocklength $\E[\tau]$ for the BSC$(0.11)$ and $\epsilon = 10^{-3}$. In this example, $k$ ranges from $1$ to $200$.}
\label{fig: rate for BSC}
\end{figure}

\begin{figure}[t]
\centering
\includegraphics[width=0.48\textwidth]{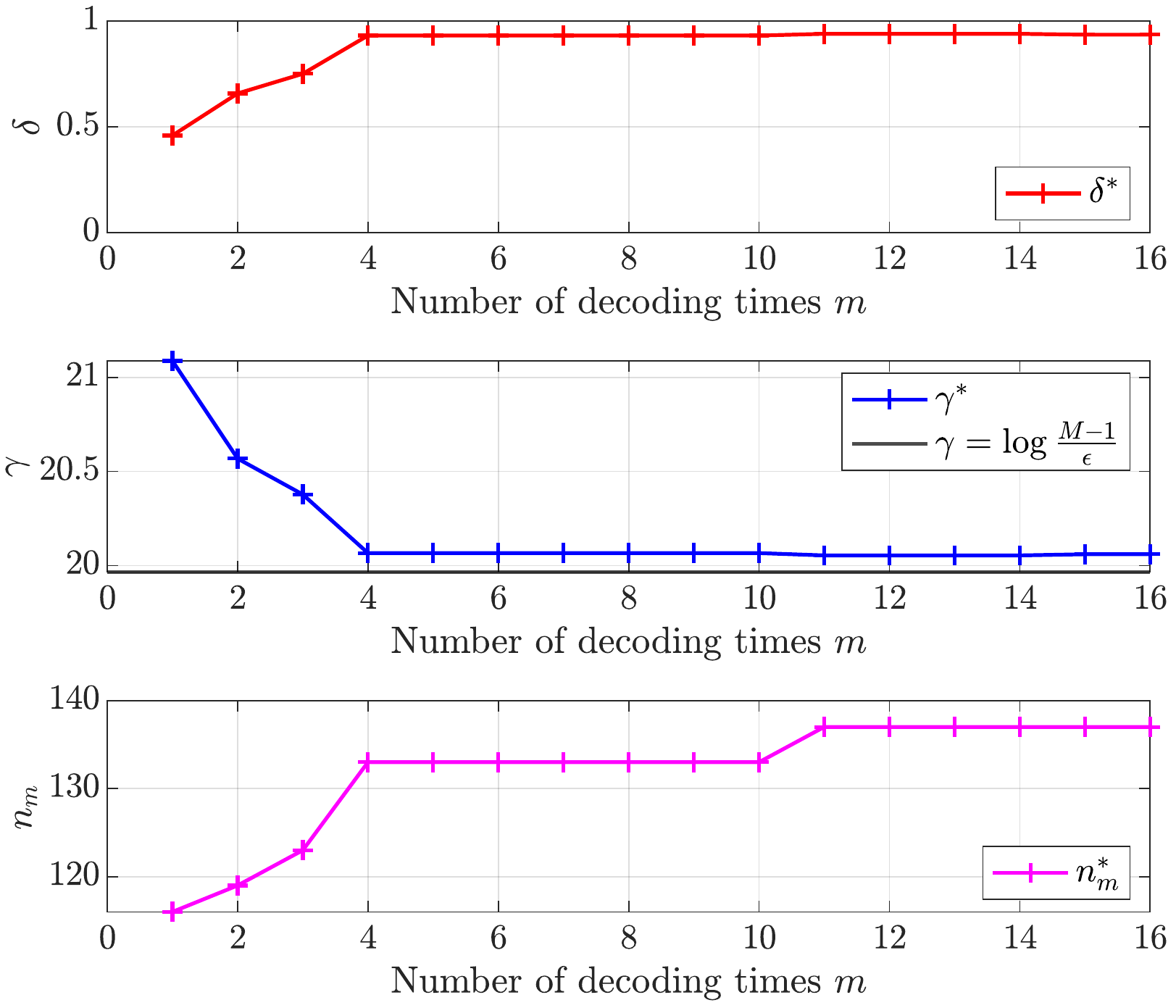}
\caption{Globally optimal $\delta^*$, $\gamma^*$, and $n_m^*$ as a function of the number of decoding times $m$ for BSC$(0.11)$ $k = 10$, and $\epsilon = 10^{-3}$. }
\label{fig: params for BSC}
\end{figure}

\begin{figure}[t]
\centering
\includegraphics[width=0.48\textwidth]{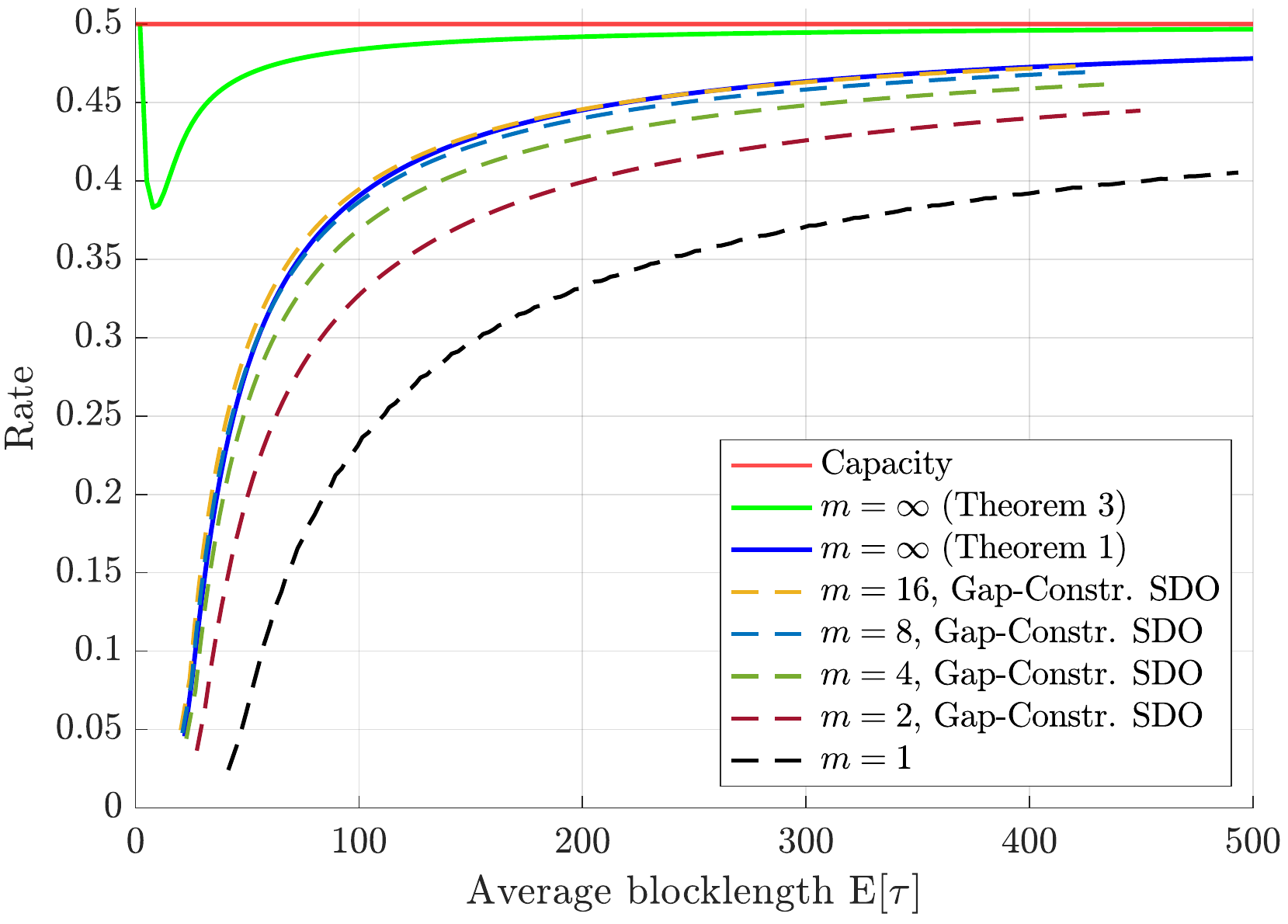}
\caption{Rate vs. average blocklength $\E[\tau]$ for the BEC$(0.5)$ and $\epsilon = 10^{-3}$. In this example, $k$ ranges from $1$ to $200$.}
\label{fig: rate for BEC}
\end{figure}

\begin{figure}[t]
\centering
\includegraphics[width=0.48\textwidth]{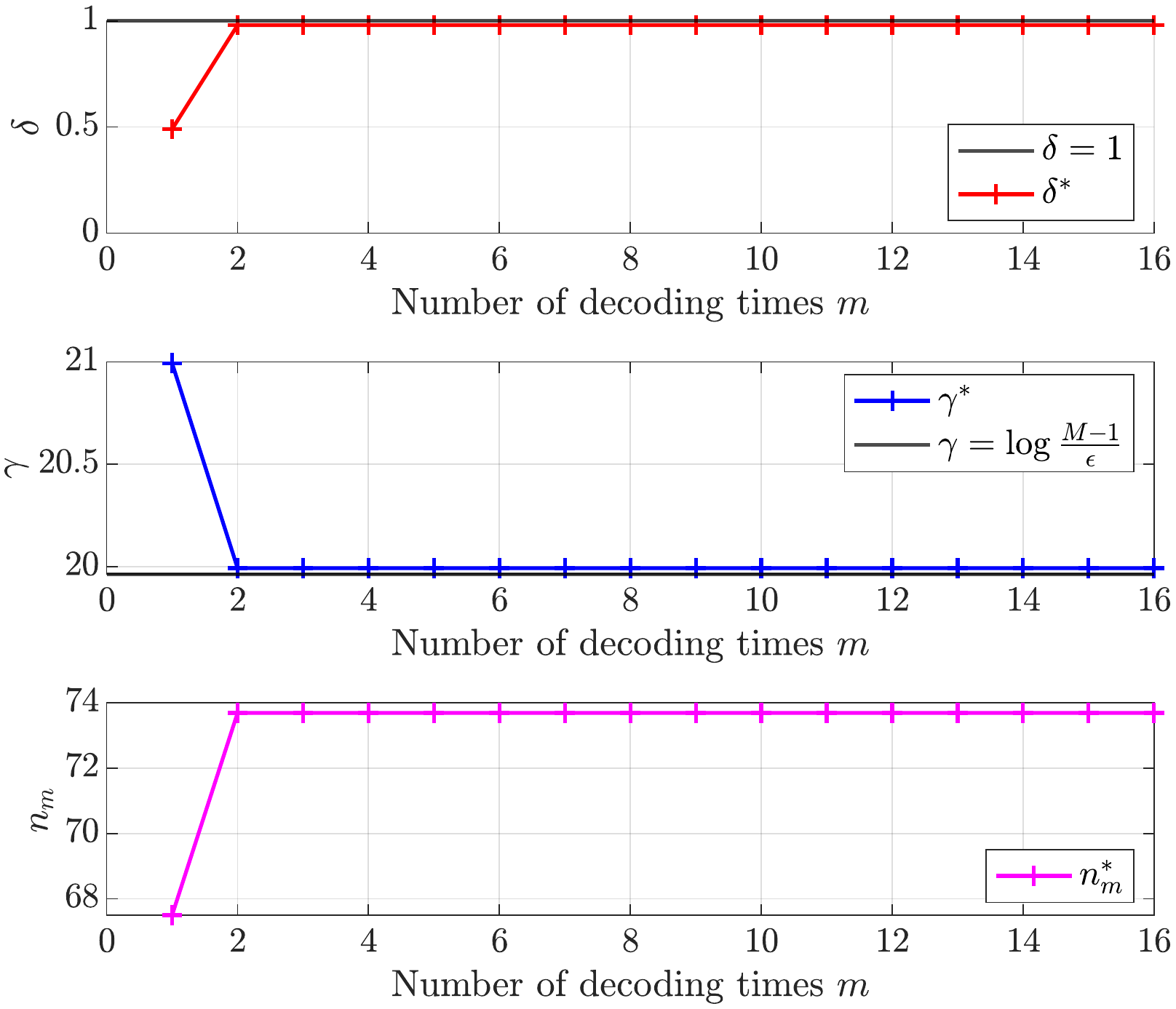}
\caption{Globally optimal $\delta^*$, $\gamma^*$, and $n_m^*$ as a function of the number of decoding times $m$ for BEC$(0.5)$ $k = 10$, and $\epsilon = 10^{-3}$. }
\label{fig: params for BEC}
\end{figure}

\subsection{BSC}

For ease of comparison with Sec. \ref{subsec: sim for BI-AWGN}, we consider the BSC with capacity $C = 0.5$ bits/channel use, which is  BSC$(0.11)$. By Theorem \ref{theorem: 6}, the approximation function $F_{\gamma}(n)$ we use to estimate $\Prob[\iota(X^n; Y^n)\ge\gamma]$ is  given by \eqref{eq: CDF of binomial}. We apply the discrete SDO procedure restricted to the set of local maximizers $\{\alpha_i\}_{i=0}^\infty$ to solve the integer program \eqref{eq: integer program}.

For $\epsilon = 10^{-3}$ and BSC$(0.11)$, Fig. \ref{fig: rate for BSC} shows achievability bounds estimated by the discrete SDO procedure and the two-step minimization, along with Polyanskiy's achievability bound. We observe a similar trend as in the BI-AWGN channel case. 
Once again, Polyanskiy's achievability bound can be approached closely with $m = 16$ for a wide range of average blocklength (or $k$).

Fig. \ref{fig: params for BSC} shows the behavior of $\delta^*$, $\gamma^*$ and $n_m^*$ as a function of the number of decoding times $m$ for $k = 10$, $\epsilon = 10^{-3}$ and BSC$(0.11)$. We observed a similar trend as in the BI-AWGN channel case. However, due to the discreteness of the information density, we see a non-smooth variation in the three parameters.


\subsection{BEC}

We consider the BEC$(0.5)$ with capacity $C = 0.5$ bits/channel use. By \eqref{eq: approximation for BEC}, we use the order-$5$ continuity-corrected Edgeworth series in Theorem \ref{theorem: corrected Edgeworth} as $F_{\gamma}(n)$, $n\in\R_+$, to approximate $\Prob[\iota(X^n; Y^n)\ge\gamma]$. We apply the gap-constrained SDO procedure along with the two-step minimization to solve the relaxed program \eqref{eq: relaxed program}.

For $\epsilon = 10^{-3}$ and BEC$(0.5)$, Fig. \ref{fig: rate for BEC} shows achievability bounds estimated by the gap-constrained SDO procedure and the two-step minimization. Previous achievability bounds for VLSF codes obtained by Polyanskiy \emph{et al.} and Devassy \emph{et al.} are also displayed. Polyanskiy's achievability bound can be closely approached or exceeded with $m = 8$ for a wide range of average blocklength (or information length $k$). With  $m = 16$, the  achievability bound estimated by the gap-constrained SDO procedure and the two-step minimization exceeds Polyanskiy's achievability bound for average blocklengths below $240$. Note that it is not surprising that Polyanskiy's achievability bound can be exceeded since the constant term is not tight as discussed in Sec. \ref{sec: preliminaries}.  

As shown by the green curve in Fig. \ref{fig: rate for BEC}, there is a significant gap between Polyanskiy's VLSF achievability bound for information density decoder and  Devassy's achievability bound for RLFC which achieves zero-error transmission (Theorem \ref{theorem: BEC achievability}).  This suggests that information density decoding is in fact a suboptimal use of the BEC.  We apply SDO to an improved version of RLFC below in  Sec. \ref{sec: ST-RLFC}.

Fig. \ref{fig: params for BEC} shows the behavior of $\delta^*$, $\gamma^*$ and $n_m^*$ as a function of the number of decoding times $m$ for $k = 10$, $\epsilon = 10^{-3}$ and BEC$(0.5)$. We see that $\delta^*$ quickly approaches $1$ as $m$ increases from $1$ to $2$, and then remains roughly constant as $m$ further increases. This trend again matches Polyanskiy's setting.


\section{VLSF Codes Under ST-RLFC for BEC}\label{sec: ST-RLFC}

Previous sections have been focused on Polyanskiy's framework of utilizing a random VLSF code and an information density decoder. However for the BEC, the decoder has the ability to identify the correct transmitted message whenever only a single codeword is compatible with the unerased received symbols. Motivated by this key observation, we propose a new random VLSF code using the systematic transmission followed by random linear fountain coding (ST-RLFC). 
The ST-RLFC scheme also facilitates a new $(l, n_1^m, 2^k, \epsilon)$ VLSF code at finite blocklength.

\subsection{The ST-RLFC Scheme}

Consider transmitting a $k$-bit message
\begin{align}
  \bm{b} = (b_1, b_2, \dots, b_k)\in\{0, 1\}^k. \label{eq: k_bit msg}
\end{align}
Let us define the set of nonzero basis vectors in $\{0, 1\}^k$ by
\begin{align}
  \G_k \triangleq \{\bm{v}\in\{0, 1\}^k:\bm{v}^\top\bm{1} > 0 \}.
\end{align}
We construct a random linear fountain code. Specifically, the channel input at time $n$ for message $\bm{b}$ is given by
\begin{align}
  X_n = \begin{cases}
    b_n, & \text{if } n\le k\\
    \bigoplus_{i=1}^k g_{n,i}b_i & \text{if } n > k,
  \end{cases} \label{eq: ST-RLFC encoder}
\end{align}
where $\oplus$ denotes bit-wise exclusive-or (XOR) operator, and $\bmg_n = (g_{n,1}, g_{n,2}, \dots, g_{n,k})^\top\in\G_k$ is generated at time $n$ according to a uniformly distributed random variable $\tilde{U}\in\G_k$ defined in Definition \ref{def: VLSF code}. Note that the encoder and decoder share the same common random variable $\tilde{U}$ at time $n > k$ so that the decoder can produce the same $\bmg_n$ at time $n$. For $1\le n \le k$, both the encoder and decoder simply use the natural basis vector $\bm{e}_n\in\R^{k\times 1}$. For all $\bm{b}\in\{0,1\}^k$, the procedure \eqref{eq: ST-RLFC encoder} specifies the common codebook before the start of transmission, i.e., the random variable $U$ in Definition \ref{def: VLSF code}.

Let $Y_n$ be the received symbol after transmitting $X_n$ over a BEC$(p)$, $p\in[0, 1)$. We consider a \emph{rank decoder} which keeps track of the rank of generator matrix $G$ associated with received symbols $Y^n$. Let $G(n)$ denote the $n$th column of $G$. If $Y_n = ?$, $G(n) = \bm{0}$; otherwise, $G(n) = \bmg_n$. Define the stopping time
\begin{align}
  \tau \triangleq \inf\{n\in\N : \text{$G(1:n)$ has rank $k$}\}, \label{eq: ST-RLFC stopping time}
\end{align}
where $G(i:j)$ denotes the column vectors from time $i$ to $j$, $1\le i\le j$. Thus, the rank decoder stops transmission at time $\tau$ and reproduces the $k$-bit message $\bm{b}$ using $Y^\tau$ and the inverse of $G(1:\tau)$. Clearly, the probability of error associated with the ST-RLFC scheme is zero.

Using ST-RLFC scheme, we obtain the a new achievability bound for zero-error VLSF codes over BEC$(p)$ in the following theorem.
\begin{theorem}\label{theorem: new BEC achiev bound}
  For a given integer $k\ge 1$, there exists an $(l, \N, 2^k, 0)$ VLSF code for BEC$(p)$, $p\in[0, 1)$, with
  \begin{align}
    l \le k + \frac{1}{C}\sum_{i=0}^{k-1}\frac{2^k-1}{2^k-2^i}F(i; k, 1-p). \label{eq: new BEC bound}
  \end{align}
  where $C = 1 - p$ and
  \begin{align}
    F(i; k, 1-p) \triangleq \sum_{j=0}^{i}\binom{k}{j}(1-p)^{j}p^{k-j} \label{eq: th13_CDF}
  \end{align}
  denotes the CDF evaluated at $i$, $0\le i\le k$, of a binomial distribution with $k$ trials and success probability $1-p$.
\end{theorem}

\begin{IEEEproof}
See Appendix \ref{appendix: proof of new bound}.
\end{IEEEproof}

For non-vanishing error probability $\epsilon > 0$, using Polyanskiy's scheme by stopping the zero-error VLSF code at $\tau = 0$ with probability $\epsilon$, the corresponding achievability bound can be readily obtained by multiplying the right-hand side (RHS) of \eqref{eq: new BEC bound} by a factor $(1 - \epsilon)$.

We remark that the new achievability bound \eqref{eq: new BEC bound} is tighter than Devassy's bound in Theorem \ref{theorem: BEC achievability}  and two bounds are equal if $p = 1$ or $k = 1$. This is stated in the following corollary.
\begin{corollary}\label{corollary: tighter bound}
  For a given $k\in\N_+$ and BEC$(p)$, $p\in[0, 1]$, it holds that
  \begin{align}
    kC + \sum_{i=0}^{k-1}\frac{2^k-1}{2^k-2^i}F(i; k, 1-p) \le k + \sum_{i=1}^{k-1}\frac{2^i-1}{2^k-2^i},
  \end{align}
  where $C = 1 - p$ and $F(i; k, 1-p)$ is given by \eqref{eq: th13_CDF}. Equality holds if $p = 1$ or $k = 1$.
\end{corollary}

\begin{figure}[t]
\centering
\includegraphics[width=0.48\textwidth]{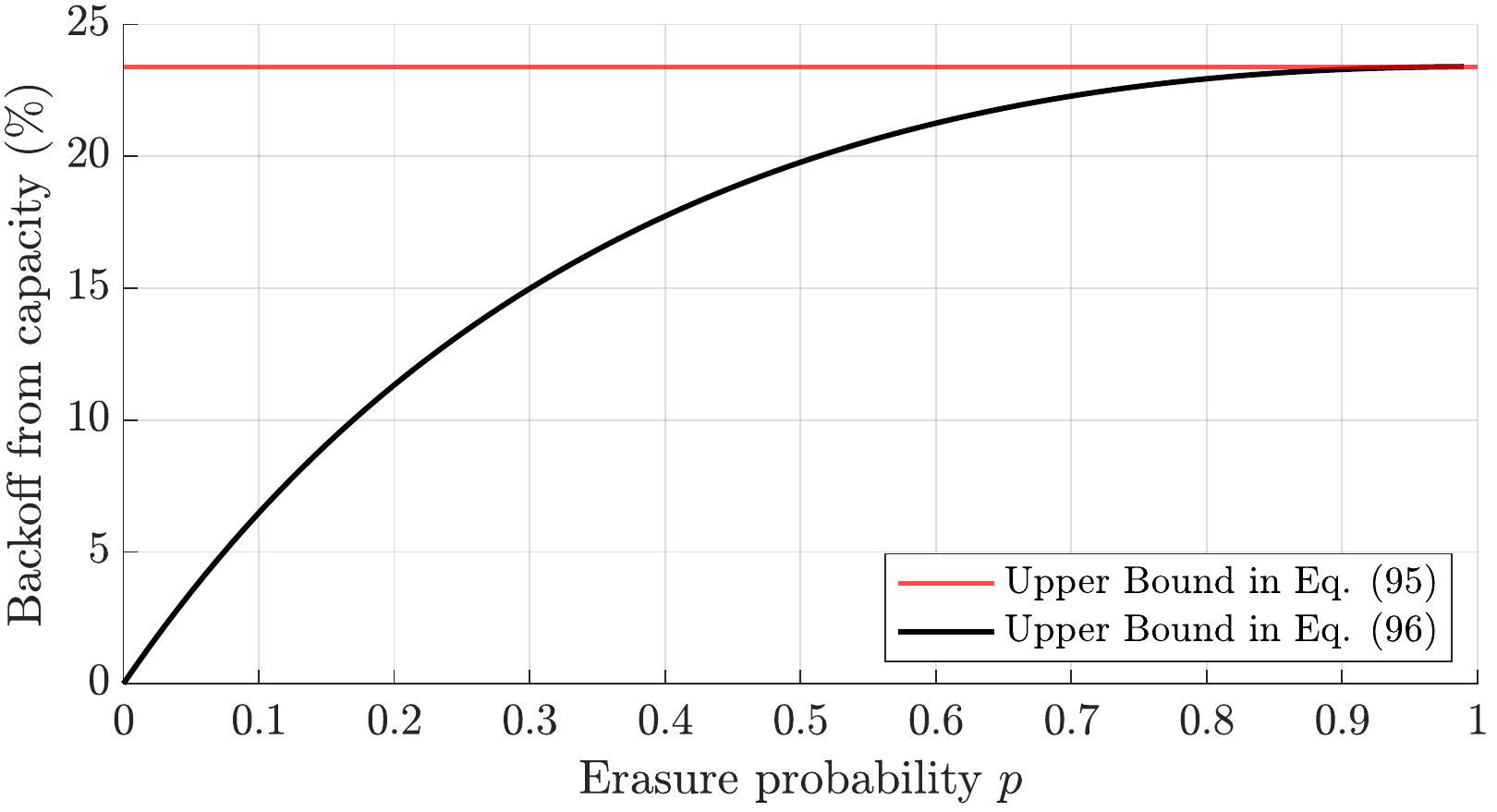}
\caption{Percentage of backoff from the capacity of BEC for $k = 3$. The red curve corresponds to a backoff percentage $23.4\%$.}
\label{fig: backoff from capacity}
\end{figure}

\begin{IEEEproof}
  Fix $k\in\N_+$ and $p\in[0, 1]$. First, note that
  \begin{align}
    k + \sum_{i=1}^{k-1}\frac{2^i-1}{2^k-2^i} = \sum_{i=0}^{k-1}\frac{2^k-1}{2^k - 2^i}.
  \end{align}
  Hence,
  \begin{align}
    &\sum_{i=0}^{k-1}\frac{2^k-1}{2^k - 2^i} - \sum_{i=0}^{k-1}\frac{2^k-1}{2^k-2^i}F(i; k, 1-p) - k(1 - p) \notag\\
    &= \sum_{i=0}^{k-1}\frac{2^k-1}{2^k-2^i}F^c(i; k, 1-p) - k(1 - p) \label{eq: cor2_eq0} \\
    &\ge \sum_{i=0}^{k-1}F^c(i; k, 1-p) - k(1 - p) \label{eq: cor2_eq1} \\
    &= 0, \notag
  \end{align}
  where in \eqref{eq: cor2_eq1}, $F^c(\cdot) \triangleq 1 - F(\cdot)$ denotes the tail probability and the sum of tail probability equals the expectation $k(1-p)$. Note that \eqref{eq: cor2_eq0} equals $0$ if $p = 1$ or $k = 1$. This completes the proof of Corollary \ref{corollary: tighter bound}.
\end{IEEEproof}
A straightforward case is BEC$(0)$ and $k\ge 2$, in which the RHS of \eqref{eq: new BEC bound} reduces to $k$, whereas the RHS of \eqref{eq: Devassy bound} is still larger than $k$. Moreover, \eqref{eq: Devassy bound} also implies an upper bound independent of $p$ on the backoff percentage from capacity,
\begin{align}
  1 - \frac{R}{C}\le \frac{\sum_{i=1}^{k-1}\frac{2^i-1}{2^k-2^i}}{k + \sum_{i=1}^{k-1}\frac{2^i-1}{2^k-2^i} }. \label{eq: old backoff percentage}
\end{align}
Devassy \emph{et al.} reported in \cite{Devassy2016} that this upper bound attains its maximum $23.4\%$ at $k = 3$, thus raising the question whether this backoff percentage is fundamental. In contrast, our result in \eqref{eq: new BEC bound} implies a refined upper bound dependent on $p$,
\begin{align}
  1 - \frac{R}{C}\le  \frac{\sum_{i=0}^{k-1}\frac{2^k-1}{2^k-2^i}F(i; k, 1-p) - kp }{\sum_{i=0}^{k-1}\frac{2^k-1}{2^k-2^i}F(i; k, 1-p) + k(1-p)}. \label{eq: new backoff percentage}
\end{align}
Fig. \ref{fig: backoff from capacity} shows the comparison of these two upper bounds at $k  = 3$. We see that for $k = 3$, the upper bound in \eqref{eq: new backoff percentage} is a strictly increasing function of $p$. As $p\to 0$, this upper bound converges to $0$, which closes the backoff from capacity at $k = 3$. As $p \to 1$, the upper bound in \eqref{eq: new backoff percentage} converges to the backoff percentage in \eqref{eq: old backoff percentage}, as shown in Corollary \ref{corollary: tighter bound}.

\subsection{New VLSF Codes With Finite Decoding Times for BECs}

The ST-RLFC scheme also facilitates a new $(l, n_1^m, 2^k, \epsilon')$ VLSF code at finite blocklength for BEC. We first present a general non-asymptotic achievability bound for such a code.
\begin{theorem}\label{theorem: new nonasymptotic achiev bound for BEC}
  Fix $n_1^m\in\N_+^m$ satisfying $n_1 < n_2 < \cdots < n_m$. For any positive integer $k\in\N_+$ and $\epsilon'\in(0, 1)$, there exists an $(l, n_1^m, 2^k, \epsilon')$ VLSF code for the BEC$(p)$ with
    \begin{align}
      l &\le n_m + \sum_{i=1}^{m-1}(n_i - n_{i+1})\Prob[S_{n_i} = k], \label{eq: nonasymptotic bound on l} \\
      \epsilon' &\le 1 - \Prob[S_{n_m} = k],  \label{eq: nonasymptotic bound on epsilon}
    \end{align}
  where the random variable $S_n$ denote the rank of the generator matrix $G(1:n)$ observed by the rank decoder. Specifically, $\Prob[S_n = k]$ is given by
  \begin{align}
    \Prob[S_n = k] = \begin{cases}
      0, & \text{if } n < k \\
      1 - \bm{\alpha}^\top T^{n-k}\bm{1}, &\text{if } n\ge k,
    \end{cases} \label{eq: expression for S_n}
  \end{align}
where $\bm{\alpha} = [\alpha_1,\alpha_2,\dots, \alpha_k]^\top\in\R^{k\times 1}$ with $\alpha_i = F(i; k, 1-p)$, $0\le i\le k-1$, where $F(i;k, 1-p)$ is given by \eqref{eq: th13_CDF}, $T\in\R^{k\times k}$ with entries given by
  \begin{align}
    T_{i, i} &= p + \frac{(1-p)(2^{i-1} - 1)}{2^k - 1}, \\
  T_{i,i+1} &= \frac{(1-p)(2^k - 2^{i-1})}{2^k - 1}, \\
  T_{i,j} &= 0, \text{ for } j\ne i \text{ and } j\ne i+1.
  \end{align}
\end{theorem}

\begin{IEEEproof}
  See Appendix \ref{appendix: proof of BEC nonasymptotic bound}.
\end{IEEEproof}

\begin{figure}[t]
\centering
\includegraphics[width=0.48\textwidth]{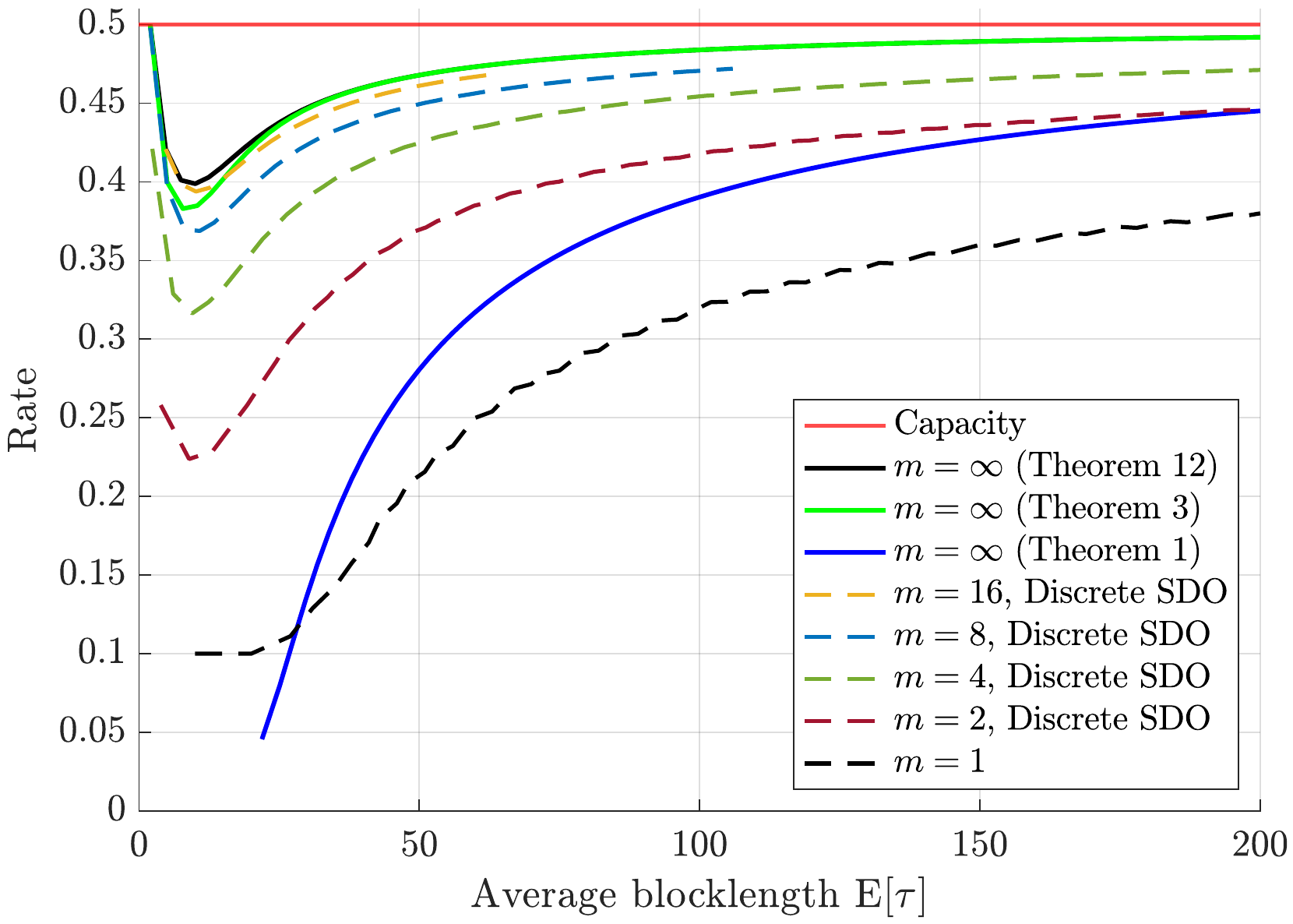}
\caption{Rate vs. average blocklength $\E[\tau]$ for the BEC$(0.5)$ and $\epsilon = 10^{-3}$ using the ST-RLFC scheme. In this example, $k$ ranges from $1$ to $100$ for $m = 1, 2, 4$; $k$ ranges from $1$ to $50$ for $m = 8$; and $k$ ranges from $2$ to $30$ for $m = 16$.}
\label{fig: rate for ST-RLFC BEC}
\end{figure}

Theorem \ref{theorem: new nonasymptotic achiev bound for BEC} facilitates a similar but a much simpler integer program. Define
\begin{align}
  N(n_1^m) &\triangleq n_m + \sum_{i=1}^{m-1}(n_i - n_{i+1})\Prob[S_{n_i} = k].
\end{align}
For a given $k\in\N_+$ and a target error probability $\epsilon\in(0,1)$, 
\begin{align}
  \begin{split}
    \min_{n_1^m}&\quad N(n_1^m) \\
    \st&\quad 1 - \Prob[S_{n_m} = k]\le \epsilon
  \end{split} \label{eq: ST-RLFC integer program}
\end{align}

Unlike the double minimization in the information density framework, the integer program \eqref{eq: ST-RLFC integer program} only involves a single minimization over $n_1^m$. Using \eqref{eq: expression for S_n}, we solve integer program \eqref{eq: ST-RLFC integer program} with the discrete SDO procedure.

Under the ST-RLFC framework, Fig. \ref{fig: rate for ST-RLFC BEC} shows the achievability bounds estimated by the discrete SDO procedure for BEC$(0.5)$ and $\epsilon = 10^{-3}$. The new achievability bound in Theorem \ref{theorem: new BEC achiev bound} along with the ones developed by Polyanskiy (Theorem \ref{theorem: 1}) and Devassy (Theorem \ref{theorem: BEC achievability}) are also shown. We see that the maximal achievable rate for $(l, n_1^m, 2^k, \epsilon)$ VLSF codes operated over a BEC$(p)$ is significantly improved, compared to the information density framework in Fig. \ref{fig: rate for BEC}. In particular, achievability bounds for $m \ge 2$  outperform Polyanskiy's achievability bound by a wide margin. The achievability bound for $m = 16$ even exceeds Devassy's bound at small values of $k$. This demonstrates that the ST-RLFC scheme further improves the VLSF code performance for BEC. 

We remark that Polyanskiy obtained a much better VLF achievability bound for the BEC \cite[Th. 7]{Polyanskiy2011} by simply retransmitting each of the $k$ bits until it gets through the BEC. However, for $k\ge 2$, this particular code construction yields a VLF code rather than a VLSF code. Therefore, the VLF achievability bound for the BEC in \cite[Th. 7]{Polyanskiy2011} is omitted from discussion.


\section{Conclusion}\label{sec: conclusion}
Practical systems use incremental redundancy with ACK/NACK feedback, but typically only have  a few decoding times that produce ACK or NACK feedback to the transmitter. In this paper, we evaluate achievability bounds for such VLSF codes with $m$ decoding times for the three classical binary-input channels. Numerical evaluations for the three channels all confirm that Polyanskiy's achievability bound, which assumes $m = \infty$, can be approached with a relatively small $m$. For example, at $\epsilon = 10^{-3}$, we show that $m = 16$ suffices to approach Polyanskiy's achievability bound. This result has the exciting implication that a variable-length codes with a small number of stop-feedback actions suffices to dramatically improve the achievable rate over that of a fixed-length code for  a given message size and target error probability. For BEC, using the ST-RLFC scheme further improves performance beyond Polyanskiy's achievability bound.

Once the approximation function $F_{\gamma}(n)$ has been determined, our techniques can identify the optimal times to attempt decoding and send feedback for any code.  What remains open is to design a deterministic VLSF code with $m$ decoding times that approaches achievability bounds demonstrated in this paper. Returning to the random-code setting, it remains open to prove that as $m\to\infty$, the achievability bound for random VLSF codes with $m$ optimal decoding times converges to Polyanskiy's result (i.e., Theorem \ref{theorem: 1}). It also remains open to develop analytical methods to understand the behavior of the optimal $\delta^*$, $\gamma^*$, and $n_m^*$ consistent with our numerical results.

\appendices

\section{Derivation of the Edgeworth Expansion}\label{appendix: Edgeworth expansion}
Our derivation is analogous to the one in \cite{Peter1992}, with the distinction that we provide explicit expression for the polynomial involved in the Edgeworth expansion.

Let $W_1, W_2, \dots, W_n$ be i.i.d. random variables with zero mean and variance $\sigma^2$. Let $\chi_W(t) = \E[e^{\ii tW}]$ be the characteristic function of $W$ and let $\{\kappa_j\}_{j=1}^\infty$ be the cumulants for $W$. Note that $\kappa_1 = 0$, $\kappa_2 = \sigma^2$.

Let $Y = W/\sigma$ be the normalized random variable. The characteristic function $\chi_Y(t) = \chi_W\Paren{t/\sigma}$. $\chi_Y(t)$ can also be expressed in terms of the exponential of a power series involving cumulants $\{\bar{\kappa}_j\}_{j=1}^\infty$, where $\bar{\kappa}_j = \sigma^{-j}\kappa_j$ denotes the $j$th cumulant of $Y$. Namely,
\begin{align}
  \chi_Y(t) = \exp\Paren{\sum_{j=1}^\infty\frac{\bar{\kappa}_j}{j!}(\ii t)^j  }, \label{eq: ap_eq0}
\end{align}
where $\bar{\kappa}_1 = 0$ and $\bar{\kappa}_2 = 1$.

Consider the standardized sum
\begin{align}
  S \triangleq \frac{1}{\sqrt{n}}\sum_{i=1}^nY_i.
\end{align}
The characteristic function $\chi_S(t)$ for $S$ is given by
\begin{align}
  \chi_S(t) = \E[\exp(\ii tS)] &= \E\Bracket{\exp\Paren{\ii  \frac{t}{\sqrt{n}}\sum_{i=1}^nY_i } } \\
    &= \Paren{\chi_Y\Paren{tn^{-\frac12 } } }^n. \label{eq: ap_eq1}
\end{align}
Substituting \eqref{eq: ap_eq0} into \eqref{eq: ap_eq1} and invoking $\bar{\kappa}_1 = 0$ and $\bar{\kappa}_2 = 1$ yields
\begin{align}
  \chi_S(t) &= \exp\Paren{\sum_{j=1}^\infty n^{-\frac{j-2}{2}}\frac{\bar{\kappa}_j}{j!}(\ii t)^j  } \notag\\
    &= \exp\Paren{-\frac12t^2 + \sum_{j=3}^\infty \frac{\bar{\kappa}_j(\ii t)^j}{j!}n^{-\frac{j-2}{2}} } \notag\\
    &= e^{-\frac{t^2}{2}}\exp\Paren{\sum_{j=1}^\infty \frac{\bar{\kappa}_{j+2}(\ii t)^{j+2}}{(j+2)!}n^{-\frac{j}{2}} } \label{eq: ap_eq3}
\end{align}
Our goal is to represent \eqref{eq: ap_eq3} as a power series. Namely,
\begin{align}
  \chi_S(t) = e^{-\frac{t^2}{2}}\Paren{1 + \sum_{j=1}^\infty n^{-\frac{j}{2}}r_j(\ii t) } \label{eq: chi_S}
\end{align}
for some polynomial $r_j(\cdot)$, $j\ge 1$. 

In \cite[Appendix A]{Blinnikov1998}, Blinnikov and Moessner proved the following useful lemma regarding the $n$th derivative of a composite function $f\circ g(x)\equiv f(g(x))$.
\begin{lemma}\label{lemma: 1}
  Let $f(x)$ and $g(x)$ be two differentiable functions with up to $n$th derivatives. Let $f^{(r)}(x)$ represent the $r$th derivative of $f(x)$ evaluated at $x$, $1\le r\le n$. Then,
  \begin{align}
    \frac{\diff^n}{\diff x^n}f(g(x)) = n!\sum_{\Brace{k_l}}f^{(r)}(y)\Big|_{y = g(x)}\prod_{l=1}^n\frac{1}{k_l!}\Paren{\frac{1}{l!}g^{(l)}(x) }^{k_l},
  \end{align}
  where $r\triangleq \sum_{l=1}^nk_l$, and the set $\{k_l\}$ consists of all non-negative integer solutions to the Diophantine equation
    \begin{align}
      k_1 + 2k_2 + \cdots + nk_n = n.
    \end{align}
\end{lemma}
As an application of Lemma \ref{lemma: 1}, with $f \equiv \exp(x)$ and $g \equiv \sum_{j=1}^\infty \frac{\bar{\kappa}_{j+2}u^{j+2}}{(j+2)!}x^{j} $, we obtain
\begin{align}
  r_j(u) &= \frac{1}{j!}\frac{\diff^j }{\diff x^j}f(g(x))\Big|_{x = 0} \notag\\
    &= \frac{1}{j!}\cdot j!\sum_{\Brace{k_i}}\prod_{i=1}^j\frac{1}{k_i!}\Paren{\frac{1}{i!}\cdot\frac{\bar{\kappa}_{i+2} u^{i+2} }{(i+2)! }i! }^{k_i} \notag\\
    &= \sum_{\Brace{k_i}}u^{j + 2r}\prod_{i=1}^j\frac{1}{k_i!}\Paren{\frac{\bar{\kappa}_{i+2}}{(i+2)!} }^{k_i}. \label{eq: ap_eq4}
\end{align}
Thus, \eqref{eq: ap_eq4} gives the polynomial $r_j(\cdot)$ that we are seeking.

Since the characteristic function for a standard normal $\phi(x)$ is exactly $e^{-t^2/2}$, the form of \eqref{eq: chi_S} suggests the following ``inverse'' expansion
\begin{align}
  \Prob[S\le x] = \Phi(x) + \sum_{j=1}^\infty n^{-\frac{j}{2}} R_j(x), \label{eq: inverse expansion}
\end{align}
where $R_j(x)$ denotes the function whose Fourier transform equals $r_j(\ii t)e^{-t^2/2}$. Our next step is to find $R_j(x)$.

Repeated integration by parts gives
\begin{align}
  e^{-t^2/2} = (-\ii t)^{-j}\int_{-\infty}^\infty e^{\ii tx}\diff \Phi^{(j)}(x), \label{eq: ap_eq5}
\end{align}
where $\Phi^{(j)}(x) = (\diff / \diff x)^j\Phi(x)$. Let $D = \diff/\diff x$ denote the differential operator. Then, \eqref{eq: ap_eq5} is equivalent to
\begin{align}
  \int_{-\infty}^\infty e^{\ii tx}\diff\Bracket{(-D)^j\Phi(x) } = (\ii t)^je^{-t^2/2}. \label{eq: ap_eq6}
\end{align}
Interpreting $r_j(-D)$ as a polynomial in $D$ so that $r_j(-D)$ itself is a differential operator. By \eqref{eq: ap_eq6}, we obtain
\begin{align}
  \int_{-\infty}^\infty e^{\ii tx}\diff\Bracket{r_j(-D)\Phi(x) } = r_j(\ii t) e^{-t^2/2}.
\end{align}
Hence, it follows that
\begin{align}
  R_j(x) = r_j(-D)\Phi(x). \label{eq: R_j poly}
\end{align}
For $j\ge 1$, we have the relation
\begin{align}
  (-D)^j\Phi(x) = -He_{j-1}(x)\phi(x), \label{eq: differential operator}
\end{align}
where $He_i(x)$ denotes the degree-$i$ Hermite polynomial, $i\ge 0$. In \cite[Eq. (13)]{Blinnikov1998}, the authors provided an explicit formula for the degree-$i$ Hermite polynomial
\begin{align}
  He_i(x) = i!\sum_{k=0}^{\lfloor i/2\rfloor}\frac{(-1)^k x^{j-2k}}{k!(j-2k)! 2^k}.
\end{align}
Combining \eqref{eq: ap_eq4}, \eqref{eq: R_j poly}, and \eqref{eq: differential operator}, we obtain
\begin{align}
  &R_j(x) = r_j(-D)\Phi(x) \notag\\
    &= \sum_{\Brace{k_i}}(-D)^{j + 2r}\Phi(x)\prod_{i=1}^j\frac{1}{k_i!}\Paren{\frac{\bar{\kappa}_{i+2}}{(i+2)!} }^{k_i} \notag\\
    &= -\sum_{\Brace{k_i}}He_{j+2r-1}(x)\phi(x) \prod_{i=1}^j\frac{1}{k_i!}\Paren{\frac{\bar{\kappa}_{i+2}}{(i+2)!} }^{k_i}. \label{eq: ap_eq7}
\end{align}
Hence, \eqref{eq: ap_eq7} gives the polynomial $R_j(x)$ we are seeking.

Finally, let us define
\begin{align}
  p_j(x) \triangleq -\sum_{\Brace{k_i}}He_{j+2r-1}(x) \prod_{i=1}^j\frac{1}{k_i!}\Paren{\frac{\bar{\kappa}_{i+2}}{(i+2)!} }^{k_i},
\end{align}
which is exactly \eqref{eq: p_j poly.}. Hence, $R_j(x) = p_j(x)\phi(x)$ for $j\ge1$. Substituting this into \eqref{eq: inverse expansion} yields
\begin{align}
  \Prob[S\le x] = \Phi(x) + \phi(x)\sum_{j=1}^\infty n^{-\frac{j}{2}} p_j(x). \label{eq: final power series}
\end{align}
In \cite{Peter1992}, it is argued that under the sufficient regularity conditions $\E[|W|^{s+2}] < \infty$, $s\in\N_+$ and $\limsup_{|t|\to\infty}|\chi_W(t)| < 1$, for all $x\in\R$, 
\begin{align}
   \Prob[S\le x] = \Phi(x) + \phi(x)\sum_{j=1}^s n^{-\frac{j}{2}} p_j(x) + o\Paren{n^{-\frac{s}{2}}} ,
\end{align}
which is exactly \eqref{eq: order-s Edgeworth expansion}. This concludes the derivation of the Edgeworth expansion.

\section{Proof of Theorem \ref{theorem: new BEC achiev bound}}\label{appendix: proof of new bound}

Let random variable $S_n$ denote the rank of generator matrix $G(1:n)$. According to the ST-RLFC scheme, the probability mass function (PMF) of $S_k$ at time $k$ is given by
\begin{align}
  \Prob[S_k = r] = \binom{k}{r}(1-p)^rp^{k-r},\quad 0\le r\le k. \label{eq: ap2_eq1}
\end{align}
For $n\ge k$, due to the BEC$(p)$ and our RLFC scheme, $S_{n+1} = S_n = r$ occurs if $Y_{n+1} = ?$ or if $Y_{n+1}\ne ?$ and $\bmg_{n+1}$ is a linear combination of previous $r$ independent basis vectors. Otherwise, $S_{n+1} = r + 1$. Hence, the behavior of $S_n$, $n\ge k$, is characterized by the following discrete-time homogeneous Markov chain with $k+1$ states.
\begin{align}
  &\Prob[S_{n+1} = r|S_n = r] = p + \frac{(1-p)(2^r - 1)}{2^k - 1}, \\
  & \Prob[S_{n+1} = r+1|S_n = r] = \frac{(1-p)(2^k - 2^r)}{2^k - 1},
\end{align}
where $0\le r\le k-1$, and $\Prob[S_{n+1} = k|S_n = k] = 1$. Note that this Markov chain has a single absorbing state $S_n = k$. The time to absorption for this Markov chain follows a discrete phase-type distribution \cite[Chapter 2]{Neuts_book}. More specifically, the one-step transfer matrix $P\in\R^{(k+1)\times (k+1)}$ of this Markov chain can be written as
\begin{align}
  P = \begin{bmatrix}
    T & \bm{t} \\
    \bm{0}^\top & 1
  \end{bmatrix}, \label{eq: ap2_one_step_transfer_matrix}
\end{align}
where the entries of $T\in\R^{k\times k}$ are given by
\begin{align}
  T_{i, i} &= p + \frac{(1-p)(2^{i-1} - 1)}{2^k - 1}, \\
  T_{i,i+1} &= \frac{(1-p)(2^k - 2^{i-1})}{2^k - 1},
\end{align}
and $T_{i,j} = 0$ for any other pair $(i, j)$, $1\le i,j\le k$. Since $P$ is a stochastic matrix, it follows that
\begin{align}
  \bm{t} = (I -T)\bm{1}.
\end{align}
The initial probability distribution is given by $[\bm{\alpha}^\top, \alpha_{k}]$, where
\begin{align}
  \bm{\alpha}^\top \triangleq \begin{bmatrix}
    \Prob[S_k=0] & \Prob[S_k=1] & \cdots & \Prob[S_k=k-1]
  \end{bmatrix}, \label{eq: initial distribution}
\end{align}
with $\Prob[S_k = r]$ given by \eqref{eq: ap2_eq1}, and $\alpha_k = 1 - \bm{\alpha}^\top\bm{1}$. Let random variable $X\in\N$ denote the time to absorbing state $k$ with initial distribution $[\bm{\alpha}^\top, \alpha_{k}]$. Hence, it follows that $X$ has PMF
\begin{align}
  \Prob[X = n] = \bm{\alpha}^\top T^{n-1}\bm{t},\quad n\in\N_+,
\end{align}
and $\Prob[X = 0] = \alpha_{k}$. Define the generating function of $X$ by
\begin{align}
  H_X(z)&\triangleq \E[z^X] = \sum_{n=0}^\infty z^n\Prob[X = n] \notag\\
    &= \alpha_k + \sum_{n=1}^{\infty}z^n\bm{\alpha}^\top T^{n-1}\bm{t} \notag\\
    &= \alpha_k + z\bm{\alpha}^\top\Paren{\sum_{n=0}^\infty (zT)^n }\bm{t} \label{eq: ap2_eq2} \\
    &= \alpha_k + z\bm{\alpha}^\top(I - zT)^{-1}(I - T)\bm{1},
\end{align}
where in \eqref{eq: ap2_eq2}, we have used $\sum_{n=0}^\infty A^n = (I - A)^{-1}$ whenever $|\lambda_i| < 1$ for all $i\in [k]$, where $\{\lambda_i\}_{i=1}^k$ denotes the eigenvalues of a square matrix $A\in\R^{k\times k}$. Hence, the expected time to absorbing state $k$ is given by
\begin{align}
  \E[X] &= \frac{\diff H_X(z)}{\diff z}\Big|_{z = 1} = \bm{\alpha}^\top(I - T)^{-1}\bm{1}.
\end{align}
Therefore, the expected stopping time $\E[\tau]$, with $\tau$ defined in \eqref{eq: ST-RLFC stopping time}, is given by
\begin{align}
  \E[\tau] &= k + \E[X] \notag\\
    &= k + \bm{\alpha}^\top(I - T)^{-1}\bm{1}  \label{eq: ap2_eq3}
\end{align}
Note that 
\begin{align}
  I - T &= (1 - p)\diag\Paren{1, \frac{2^k-2^1}{2^k-1},\frac{2^k-2^2}{2^k-1},\cdots, \frac{2^k-2^{k-1}}{2^k-1} } \notag\\
    &\phantom{==}\cdot\begin{bmatrix}
      1 & -1 & 0 & \cdots & 0 \\
      0 & 1 & -1 & \cdots & 0 \\
      0 & 0 & 1 & \cdots & 0 \\
      \vdots & \vdots & \vdots & \ddots & \vdots \\
      0 & 0 & 0 & \cdots & 1
    \end{bmatrix}
\end{align}
Hence, 
\begin{align}
  &(I - T)^{-1} = (1-p)^{-1}\begin{bmatrix}
      1 & 1 & 1 & \cdots & 1 \\
      0 & 1 & 1 & \cdots & 1 \\
      0 & 0 & 1 & \cdots & 1 \\
      \vdots & \vdots & \vdots & \ddots & \vdots \\
      0 & 0 & 0 & \cdots & 1
    \end{bmatrix}\notag\\
    &\phantom{==}\cdot\diag\Paren{1, \frac{2^k-1}{2^k-2^1}, \frac{2^k-1}{2^k-2^2},\cdots \frac{2^k-1}{2^k - 2^{k-1}} } \notag\\
  &= (1-p)^{-1}\begin{bmatrix}
      1 & \frac{2^k-1}{2^k-2^1} & \frac{2^k-1}{2^k-2^2} & \cdots & \frac{2^k-1}{2^k - 2^{k-1}} \\
      0 & \frac{2^k-1}{2^k-2^1} & \frac{2^k-1}{2^k-2^2} & \cdots & \frac{2^k-1}{2^k - 2^{k-1}} \\
      0 & 0 & \frac{2^k-1}{2^k-2^2}  & \cdots & \frac{2^k-1}{2^k - 2^{k-1}} \\
      \vdots & \vdots & \vdots & \ddots & \vdots \\
      0 & 0 & 0 & \cdots & \frac{2^k-1}{2^k - 2^{k-1}}
    \end{bmatrix}. \label{eq: inverse matrix}
\end{align}
Substituting \eqref{eq: initial distribution} and \eqref{eq: inverse matrix} into \eqref{eq: ap2_eq3}, we finally obtain
\begin{align}
  \E[\tau] &= k + (1-p)^{-1}\sum_{i=0}^{k-1}\frac{2^k-1}{2^k - 2^i}\sum_{j=0}^i\Prob[S_k = j] \\
    &= k + \frac{1}{C}\sum_{i=0}^{k-1}\frac{2^k-1}{2^k - 2^i}F(i; k, 1-p), \label{eq: ap2_eq4}
\end{align}
where $C = 1 - p$ and $F(i;k,1-p) \triangleq \sum_{j=0}^i\Prob[S_k = j]$ denotes the CDF evaluated at $i$ of a binomial distribution with $k$ trials and success probability $1 - p$. Since \eqref{eq: ap2_eq4} is the expected stopping time for an ensemble of zero-error VLSF codes, there exists an $(l, \N, 2^k, 0)$ VLSF code with
\begin{align}
  l \le k + \frac{1}{C}\sum_{i=0}^{k-1}\frac{2^k-1}{2^k - 2^i}F(i; k, 1-p).
\end{align}
This concludes the proof of Theorem \ref{theorem: new BEC achiev bound}.

\section{Proof of Theorem \ref{theorem: new nonasymptotic achiev bound for BEC}} \label{appendix: proof of BEC nonasymptotic bound}

The proof builds upon the proof of Theorem \ref{theorem: new BEC achiev bound} with the distinction that we need to specify the rank decoder for a given set of decoding times $n_1, n_2, \dots, n_m$.

Fix $n_1^m\in\N_+$ with $n_1 < n_2 < \cdots < n_m$. For a given $k\in\N_+$ and $\epsilon'\in(0, 1)$, the encoder of a random $(l, n_1^m, 2^k, \epsilon')$ VLSF code is the same as described in \eqref{eq: ST-RLFC encoder}. The rank decoder still shares the same common randomness with the encoder in selecting the basis vector $\bmg_n$, except that it now adopts the following stopping time:
\begin{align}
  \tau^* \triangleq \inf\{n\in\{n_i\}_{i=1}^m: G(1:n) \text{ has rank $k$ or $n = n_m$} \}.
\end{align}
If $\tau \le n_m$ and $G(1:\tau)$ is full rank, the rank decoder reproduces the transmitted message using $Y^\tau$ and the inverse of $G(1:\tau)$. If $\tau = n_m$ and $G(1:n_m)$ is rank deficient, then the rank decoder outputs an arbitrary message.

Let $S_n$ denote the rank of the generator matrix $G(1:n)$ observed at the rank decoder. The expected stopping time $\E[\tau^*]$ is written as
\begin{align}
  \E[\tau^*] &= \sum_{n=0}^\infty\Prob[\tau^* > n] \notag\\
    &= n_1 + \sum_{i=1}^{m-1}(n_{i+1} - n_i)\Prob[\tau^* > n_i] \\
    &= n_1 + \sum_{i=1}^{m-1}(n_{i+1} - n_i)\Prob[S_{n_i} < k] \\
    &= n_m + \sum_{i=1}^{m-1}(n_{i} - n_{i+1})\Prob[S_{n_i} = k],
\end{align}
which is equal to the upper bound in \eqref{eq: nonasymptotic bound on l}.

Note that at finite blocklength, the error only occurs when the rank of generator matrix $G(1:n_m)$ is still less than $k$. Hence,
\begin{align}
  \epsilon' &\le \Prob[S_{n_m} < k] \\
    &= 1 - \Prob[S_{n_m} = k],
\end{align}
which is equal to the upper bound in \eqref{eq: nonasymptotic bound on epsilon}. 

At time $n < k$, due to the systematic transmission, $\Prob[S_n = k] = 0$. At time $n\ge k$, as discussed in Appendix \ref{appendix: proof of new bound}, the behavior of $S_n$ is characterized by a discrete-time homogeneous Markov chain with $k+1$ states whose one-step transfer matrix is given by \eqref{eq: ap2_one_step_transfer_matrix}, and whose initial probability distribution is $[\bm{\alpha}^\top, \alpha_k]$, where $\bm{\alpha}^\top$ is given by \eqref{eq: initial distribution}. Hence, for $n \ge k$,
\begin{align}
  \Prob[S_n = k] &= 1 - \Prob[S_n < k] \\
    &= 1 - \bm{\alpha}^\top T^{n-k}\bm{1}.
\end{align}
This completes the proof of Theorem \ref{theorem: new nonasymptotic achiev bound for BEC}.


%





\ifCLASSOPTIONcaptionsoff
  \newpage
\fi

\bibliographystyle{IEEEtran}
\bibliography{IEEEabrv,references}
\end{document}